\documentclass{aa}
\usepackage{natbib}
\usepackage[]{graphicx}
\usepackage{txfonts}
\usepackage{lscape}
 \begin{document} 

\title{Near-infrared polarization images of the Orion proplyds}

 \author{S. Rost\inst{1} \and A. Eckart\inst{1,2} \and T. Ott\inst{3}}
              
 \institute{I. Physikalisches Institut, University of Cologne, Z\"ulpicher Str. 77, 50937 Cologne, Germany\\
 \email{rost@ph1.uni-koeln.de}
\and Max-Planck-Institut f\"ur Radioastronomie, Auf dem H\"ugel 69, 53121 Bonn, Germany
 \and  Max-Planck-Institut f\"ur extraterrestrische Physik, Giessenbachstr., 85748 Garching, Germany}

  \date{Received March 27, 2007; accepted March 27, 2008}

\abstract{}{The aim is to study structure and polarization properties of the stars and planet systems in the active Orion \ion{H}{II}-region.}
{We performed AO-assisted high-resolution imaging polarimetry on selected Orion proplyds close to the 
Trapezium stars in the J, H, and K bands. Differential polarimetric images of one of the largest and brightest proplyds 
are interpreted using
3D radiation transfer simulations based on the Monte Carlo method.}
{Although not fully resolvable by ground-based observations, the circumstellar material can be mapped with polarimetry.
We present constraints on
the disk parameters of the giant proplyd 177-341. 
We tested whether dust models with different grain size distributions could explain 
the observed extent of the polarization patterns and find that  
simple models with larger grains will not reproduce the spectral polarization behavior.}
{The technique of polarimetric differential imaging (PDI) in the NIR provides a good opportunity to study the structure
of the Orion proplyds.}



 \maketitle 

\section{Introduction}
\label{sect:intro}  

The Orion molecular clouds at a distance of about 460 
parsecs are the nearest location of ongoing massive star 
formation. This region also harbors one of the densest young 
stellar clusters \citep{1994AJ....108.1382M}, the Orion Nebula Cluster (ONC).  The core of the 
ONC is known as the Trapezium cluster, named after the four massive 
OB stars at its very center. The dominant Trapezium star is
$\theta^1$ Ori C, which 
generates most of the intense ultraviolet radiation field
that photo-dissociates and photo-ionizes the surrounding material.
The discovery of proto-planetary disks (proplyds) in this region
by \cite{1993ApJ...410..696O} with the Hubble Space Telescope (HST) has 
provided an unprecedented opportunity to study planet-forming 
disks around young stars. 
These objects are surrounded by bright, crescent, and tear-drop 
shaped ionization fronts (IF) resulting from gas flow off of the 
disk. At greater distances and shielded from radiation, several 
proplyds without ionization front have been observed. They are
seen in silhouette against 
the nebular background \citep{2000AJ....119.2919B}. 

New insights into the nature of proplyds have been acquired 
through high-resolution multi-spectral imaging with HST 
\citep{1998AJ....116..293B},
follow-up imaging, and spectroscopy 
at a variety of wavelengths from the ground, and intensive 
theoretical modeling \citep[e.g.]{1998ApJ...499..758J,1999ApJ...515..669S}. 
As a result, an 
irradiated disk model has been developed to account 
for the presently known observed features.

Near infrared (NIR) observations of the Trapezium cluster 
\citep{1998AJ....116.1816H,2004AJ....128.1254L} 
have shown that most young stars have excess emission 
in this wavelength regime. This is interpreted as the presence of 
circumstellar disks. According to the study by \cite{2000AJ....120.3162L},
80\% of the stars in the Trapezium cluster
show an infrared excess and are likely to be surrounded by disks.

Currently much activity is devoted to obtain physical properties of the circumstellar
dust distribution and these disks. \cite{2006ApJ...642.1140O} showed 
that encounters of stars and circumstellar disks in the 
Trapezium cluster are quite frequent and thus can cause a 
considerable contribution to the mass loss and the truncation 
of proplyds. 
Encounter-triggered planet formation is also possible as an 
important supplement to triggering by super massive stars
\citep{2003ASPC..287..263B} and supernovae-triggering
of the OB-stars \citep{2005ASPC..341..107H}. The disks of the proplyds close
to the Trapezium with a typical radius of 20--80 AU are not resolvable directly.
In the Orion nebula the YSO 177-341, also known as HST1 is one of the best-studied proplyds because
of its large size and orientation \citep{1998AJ....116..293B,1999AJ....118.2350H,2000RMxAC...9..198H}. 
The very bright ionization front outshines details of its inner structure. 

\cite{2006A&A...446..201M} showed that polarimetry 
can be used to trace the relative orientation of disks in young binary systems.
The polarization structure of a massive young
stellar object (YSO) in OMC-1 revealed a surrounding disk/bipolar outflow 
system \citep{2005Natur.437..112J}. 
Near-infrared polarimetry is a proven remedy for mapping hidden effects and 
structures \citep[see, for example,][]{1997A&A...326..632A,2006A&A...460...15M}. 
In the case of dust around a bright star, only the scattered light from the disk is expected to be polarized,
the unpolarized light from the central star is suppressed by imaging polarimetry. Therefore this technique can
improve the contrast and the detection-threshold, and it provides a possibility to map structures very close to bright stars
without coronographic masks \citep{2001ApJ...553L.189K,2006MNRAS.365.1348H}.
Polarimetric images can also trace the sources of illumination of the reflection nebula in star-forming regions \citep{2007PASJ...59..221H}.

We carried out a JHK polarimetry survey of selected proplyds in the Trapezium Cluster to study dust and disk parameters in
this dense star-forming region. Those disks of the proplyds close to the Trapezium that have a bright IF are so far only seen as
IR excesses. 
With the polarimetric images we want to map the circumstellar 
dust distribution and extract parameter ranges for 
the disks to classify the Trapezium proplyds. 
In this paper we want to present the results from our first polarimetric study of the Orion proplyds. We begin in Sect. 2 with 
the description of the observational data and summarize in Sect. 3 the basic properties of one of the sources extracted from the 
polarized data set. In the next section we consider different analytical techniques for a more detailed view of the proplyds properties 
including modeling of circumstellar dust particles and qualitative comparison 
to simulated NIR polarimetric images. A summary and an outlook
of forthcoming surveys is noted in Sect. 5. 

\section{Observations and data reduction}
\label{chap:obs}
\begin{figure}
\centering
\includegraphics[width=0.45\textwidth]{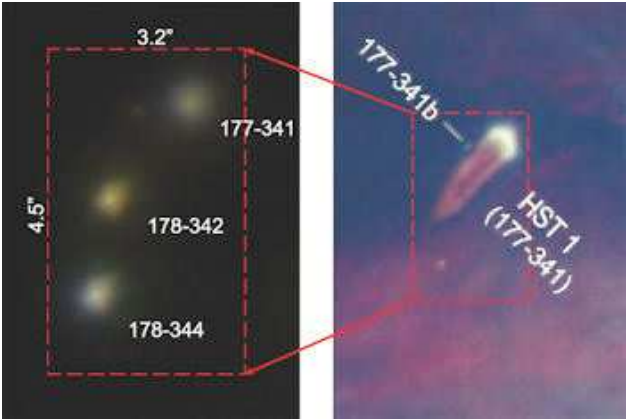}
 \caption{Near-Infrared (left, KHJ $\equiv$ RGB) image of the giant proplyd 177-341 
and its neighbors in comparison to the optical HST image 
\citep[right, red is \lbrack\ion{N}{II}\rbrack, green is \ion{H}{$_\alpha$}, blue is \lbrack\ion{O}{I}\rbrack,][]{1998AJ....116..293B},
with north facing upwards and east to the left. The proplyd 178-342 is covered by the ionization front in the optical, but barely visible at the end
of the tail.}
\label{ifoptn}
\end{figure} 

High-resolution near-infrared polarimetric images were acquired in 2001 with the ESO Adaptive Optics
System ADONIS and the SHARP II+ camera at the La Silla 3.6m\footnote{ESO program-ID: 066.C-0219} telescope \citep{1999A&AS..138..163A}.
A rotating wire grid polarizer provided polarization angles from $0^\circ$ to $150^\circ$
with a $30^\circ$ increment. 
The sources are 10 proplyds close to the Trapezium stars, embedded in three pointings with a 
field-of-view (FOV) of 8.96''$\times$8.96'' and a 0.035'' plate scale. Images were taken  
over three nights in the J, H, and K band.
No occulting masks were used in these observations due to multiple sources per FOV.
The chosen detector integration time for a good signal-to-noise ratio (SNR) and 
dynamic range was 15 s 
per polarization angle and FOV. The total time
spent on each source was 6750 s, with several measurements per angle in different nights 
(375 s per source, angle, and band).
$\theta^1$ Ori C was used for wavefront sensing
of the adaptive optics. The natural seeing varied from 0.5'' to 1''.
The achieved Strehl ratio in JHK was 6\%, 13\%, and 20\%
for the most distant FOV.
Off-source sky for each angle and dark frames were taken, 
as well as frames with unpolarized reference stars.

The data reduction was carried out using the DPUSER 
software for astronomical image analysis \citep[T. Ott, 
http://www.mpe.mpg.de/$\sim$ott/dpuser/, see also][]{1990ASPC...14..336E} and IDL routines. 
Due to some cooling problems of the detector, 
some part of the data had 
to be sorted out by a statistical filter algorithm followed by a manual inspection. 
The raw images were sky-subtracted and flat field-corrected. 

The non-polarized light consists mainly of the central star`s light and the uncorrected AO residuals,
which is the dominant noise source of ground-based AO-observations in the NIR.
Polarimetric images are very sensitive to minor deviations of
shift-and-add operations. Therefore this reduction step was carefully
carried out and manually optimized for every single frame. 
A minimal number of three frames (spanning more than $90^\circ$ position angle) is required to determine
the linear polarization and its position angles.
The redundant data set of three orthogonal polarization component pairs 
increases the SNR and reduces the instrumental polarization. 
By fitting the oversampled polarization curve with a cosine function, the linear polarization 
angle and degree maps ($\Pi_L$) have been determined. 
Polarimetric differential images (PDI) can be obtained
using $Q_0=I_{0^\circ}-I_{90^\circ}$ (Stokes Q-parameter), $Q_{30}=I_{30^\circ}-I_{120^\circ}$, and $Q_{60}=I_{60^\circ}-I_{150^\circ}$. 
In the widely used notation of linear polarized light this converts to $Q=Q_0$, and
$$U=\frac{Q_{\phi}-Q\cdot\cos(2\phi)}{\sin(2\phi)}$$ with $\phi=30^\circ$ or $\phi=60^\circ$, 
and $$\Pi_L=\frac{\sqrt{Q^2+U^2}}{I}, \; \theta=\frac{1}{2} \arctan{\frac{U}{Q}}\;.$$
The Q- and U-parameters are used for better comparison to the simulated images of the model.
\section{Results}

Some of the sources show a clearly extended structure in the NIR referring to an illuminated dust envelope.
The surrounding nebula emission is not
as strong and spatially varying as in the optical.  
The properties of the targets are listed in Table \ref{tabprop}, including the IF-sizes and NIR magnitudes, J-H
and H-K colors.
These were photometrically calibrated with the standard star HD98161 from the 2MASS catalog.
This unpolarized star was also used for polarimetric calibration. 
The aperture polarization of the targets is shown in Table \ref{tabpol}. Also the integrated polarization in two annuli is shown
for comparison and classification of circumstellar material.
The inner region is usually dominated 
by alignment and subtraction residuals \citep{2006MNRAS.365.1348H}. The annulus of 0.3''-0.5'' refers to the circumstellar dust,
which is illuminated by the central star. The outer annulus should be dominated by polarized light from the surrounding reflecting nebula where
the source of illumination is unknown.

\begin{figure}
\centering
\includegraphics[height=0.48\textwidth,angle=90]{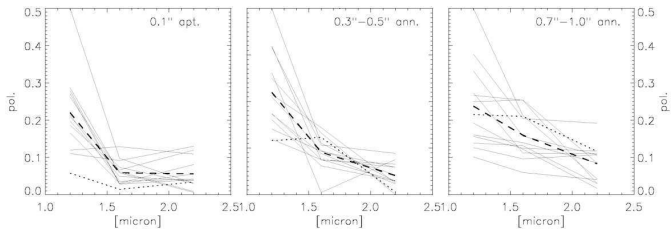}
\caption{Spectral distribution of polarization for J, H, and K bands. In an aperture of 0.1'' the light of the central star
the scattering by foreground material dominates. In an annulus close to the star the polarization is higher due to the circumstellar envelope and disk.
The outer annulus is dominated by the reflection nebula and is highly polarized. 
The dashed line is the average polarization spectrum,
the dotted line is 177-341.
}
\label{jhkplot}
\end{figure}

The polarization values as a function of the wavelength are plotted in Fig. \ref{jhkplot}.
Although the integrated polarization degree and angle is highly affected by minor deviations in shift-and-add operations and PSF
effects (see Sect. \ref{simchap}), the average values give a good overview of the circumstellar polarization distribution.
The nebula emission in the outer annuli is polarized from 10\% in K-band to 20\% in J-band. This is in good agreement with
a recent wide-field polarization study of the ONC in the NIR \citep{2006ApJ...649L..29T}.
 
\begin{figure}
\centering
\includegraphics[width=0.48\textwidth]{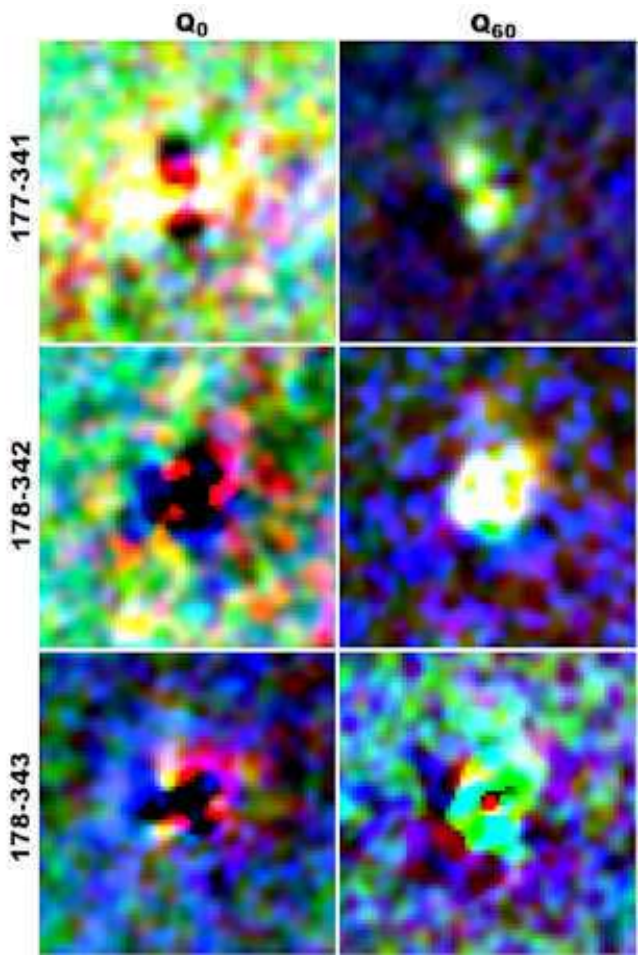}
\caption{Polarimetric differential images (KHJ $\equiv$ RGB) of the sources in the FOV shown in Fig. \ref{ifoptn}. 
Top: Proplyd 177-341 in $Q_0$ (left)
and $Q_{60}$ (right). A clear butterfly pattern is visible in all bands. Mid and Bottom: 178-342 and 178-343 show a 
trefoil that is supposed to be an AO-effect. With this comparison of sources in one FOV, lens- (e.g. astigmatism) and
PSF-effects can be excluded as the origin for the results of 177-341.}
\label{resrgb}
\end{figure}

To remove all non-polarized light from the sources, we
subtracted orthogonal components from each other.
These images of the Stokes Q- or U-parameters show a characteristic trefoil pattern for most
of the sources (Fig. \ref{resrgb}). 
The trefoil is a mode in the AO wavefront reconstruction that may contain residual power if the AO-loop is run under
variable atmospheric conditions.
However, for the giant proplyd 177-341
we find significant butterfly patterns or two lobes, which is the polarization pattern
expected for a circumstellar
dust envelope.  

The sources around 177-341 (Fig. \ref{resrgb}) lie in
a relatively dark region of the nebula.  
This is why we concentrate here in particular on the field of the proplyds around 177-341.
The NIR-image in Figure \ref{ifoptn} shows quite a 
different view of the proplyds morphology than the known optical 
images \citep{1998AJ....116..293B}. First of all the bright ionization front,
which covers inner structures of the proplyds in the optical is not visible in the NIR-images.
This reveals, for example, a previously unknown source behind the tail of the proplyd 177-341 (Fig. \ref{ifoptn}, left).
If the new source is a proplyd, it would be named 178-342, following the designation 
system introduced by \cite{1994ApJ...436..194O}. 
The observed JHK-colors differ from the neighbor 177-341, 178-342 
seems redder (Table \ref{tabprop}). 
This may be due to extinction by the overlying ionization front.

\begin{table*}
\caption{JHK-magnitudes and colors.}
\label{tabprop}
\centering  
 \begin{tabular}{lccccccccc}
 \hline\hline
Designation & R.A. & DECL. &  Dist. to& IF$^{\mathrm{a}}$ (opt.) & \multicolumn{3}{c}{Magnitude$^{\mathrm{b}}$} &  \multicolumn{2}{c}{Color}   \\
& (J2000) & (J2000) & $\theta^1$ Ori C & Tail & J & H & K$_S$ & J-H & H-K  \\ \hline

157-323 & 05 35 15.71 & -05 23 22.59 &  11.25'' & $<$0.7'' & 9.88&9.05&8.93&0.83&0.11 \\
158-323 & 05 35 15.82 & -05 23 22.50 &  9.75'' & 1.8--2.5'' & 9.73&8.83&8.53&0.91&0.3  \\
158-326 & 05 35 15.83 & -05 23 25.62 &  10.2'' & 0.85'' & 12.9&11.34&10.65&1.56&0.69 \\
158-327 & 05 35 15.79 & -05 23 26.61 &  10.53'' & 1.0--1.5'' & 11.68&10.4&9.66&1.28&0.75 \\
163-317 & 05 35 16.27 & -05 23 16.72 & 6.42'' & 1.6'' & 10.14&9.41&9.2&0.72&0.21 \\
166-316 & 05 35 16.60 & -05 23 16.32 &  7.36'' & 0.4'' & 11.4&10.47&10.2&0.93&0.27 \\
167-317 & 05 35 16.73 & -05 23 16.63 &  7.08'' & 0.67''$\times$2.1'' & 10.37&9.41&8.94&0.96&0.47 \\
177-341 & 05 35 17.67 & -05 23 40.96 &  25.98'' & 0.8''$\times$3.5'' & 12.44&11.83&11.53&0.61&0.3 \\
178-342 & 05 35 17.76 & -05 23 42.50 &  27.2'' & $\ldots$ & 12.93&11.67&11.2&1.26&0.46 \\
178-344 & 05 35 17.78 & -05 23 44.25 &  29.1'' & $\ldots$ & 12.083&11.28&11.1&0.8&0.17 \\
HD98161 & 11 17 12.01 & -38 00 51.72 & $\ldots$ & $\ldots$& 6.075 & 6.019 & 5.992 & 0.056 &0.027 \\
 \hline
  \end{tabular}
\begin{list}{}{}
\item[$^{\mathrm{a}}$] \cite{1998AJ....116..293B}
\item[$^{\mathrm{b}}$] magnitude errors $\sim0.1$
\end{list}
\end{table*}

\begin{table*}
\caption{Integrated polarization of the observed proplyds and surrounding reflection nebula in an 
aperture of  0.1'', in an annulus of 0.3''--0.5'' and 0.7''--1.0'', respectively. }
\label{tabpol}
\centering
 \begin{tabular}{llrlrlrlrlrlrl}
 \hline\hline
Designation & Filter & \multicolumn{4}{c}{Aperture 0.1
"} & \multicolumn{4}{c}{Annulus 0.3"-0.5"} & \multicolumn{4}{c}{Annulus 0.7"-
1.0"} \\ 
& &  \multicolumn{2}{c}{Pol.[\%]} &  \multicolumn{2}{c}{Ang.[$^\circ$]} &   \multicolumn{2}{c}{Pol.[\%]} &  \multicolumn{2}{c}{Ang.[$^\circ$]}
&   \multicolumn{2}{c}{Pol.[\%]} &  \multicolumn{2}{c}{Ang.[$^\circ$]}\\ \hline
157-323 & J & 18.11&$\pm$0.70 & 35.5&$\pm$1.0 & 14.38&$\pm$0.25 & 45.5&$\pm$0.3 & 
14.89&$\pm$0.37 & 38.7&$\pm$0.8 \\ 
 & H & 4.76&$\pm$0.26 & 84.4&$\pm$5.0 & 6.41&$\pm$1.96 & 22.7&$\pm$0.2 & 
11.95&$\pm$0.68 & 26.0&$\pm$1.0 \\ 
 & K$_S$ & 3.71&$\pm$1.91 & 62.1&$\pm$4.7 & 2.77&$\pm$0.64 & -67.2&$\pm$29.6
 & 11.25&$\pm$0.71 & -56.7&$\pm$0.9 \\ 
158-323 & J & 19.61&$\pm$2.09 & 21.0&$\pm$2.1 & 13.33&$\pm$1.65 & 44.6&$\pm$0.3 & 
12.11&$\pm$0.77 & 40.3&$\pm$0.5 \\ 
 & H & 3.58&$\pm$0.18 & 61.3&$\pm$2.7 & 5.33&$\pm$2.22 & 20.8&$\pm$1.8 & 9.30
&$\pm$0.63 & 23.3&$\pm$0.4 \\ 
 & K$_S$ & 4.12&$\pm$1.22 & 24.4&$\pm$6.0 & 2.72&$\pm$0.92 & -60.4&$\pm$31.4
 & 10.31&$\pm$0.37 & -51.5&$\pm$0.6 \\ 
158-326 & J & 45.32&$\pm$4.52 & 35.9&$\pm$2.1 & 32.53&$\pm$0.71 & 29.9&$\pm$0.7 & 
46.87&$\pm$1.53 & 29.4&$\pm$0.1 \\ 
 & H & 9.12&$\pm$2.50 & 89.2&$\pm$1.6 & 9.26&$\pm$1.31 & 29.9&$\pm$0.6 & 
20.06&$\pm$0.74 & 29.1&$\pm$0.1 \\ 
 & K$_S$ & 3.08&$\pm$1.17 & -76.4&$\pm$10.5 & 2.61&$\pm$0.46 & 42.2&$\pm$7.4
 & 4.30&$\pm$0.61 & 17.6&$\pm$2.0 \\ 
158-327 & J & 19.14&$\pm$3.42 & 32.8&$\pm$0.4 & 17.27&$\pm$0.93 & 40.9&$\pm$1.1 & 
18.24&$\pm$0.85 & 35.0&$\pm$0.2 \\ 
 & H & 6.94&$\pm$1.69 & -78.4&$\pm$3.4 & 7.95&$\pm$1.25 & 34.0&$\pm$0.8 & 
11.69&$\pm$0.21 & 33.7&$\pm$0.3 \\ 
 & K$_S$ & 3.07&$\pm$1.84 & -40.2&$\pm$16.0 & 3.76&$\pm$0.42 & 58.1&$\pm$7.4
 & 1.93&$\pm$0.22 & -88.7&$\pm$2.3 \\ 
163-317 & J & 24.77&$\pm$2.72 & 34.9&$\pm$1.5 & 9.92&$\pm$0.44 & 59.5&$\pm$2.9 & 
13.23&$\pm$1.65 & 71.0&$\pm$0.3 \\ 
& H & 2.89&$\pm$1.66 & -29.5&$\pm$23.1 & 6.41&$\pm$1.91 & 68.3&$\pm$4.3 & 
12.57&$\pm$0.87 & -87.3&$\pm$0.3 \\ 
& K$_S$ & 3.96&$\pm$1.61 & -79.1&$\pm$5.5 & 5.10&$\pm$0.98 & 29.9&$\pm$1.0
 & 3.12&$\pm$0.43 & -24.8&$\pm$21.0 \\ 
166-316 & J & 25.50&$\pm$1.58 & 44.3&$\pm$1.3 & 14.28&$\pm$1.95 & 54.8&$\pm$1.7 & 
9.75&$\pm$1.33 & 77.5&$\pm$0.5 \\ 
 & H & 3.00&$\pm$0.94 & 36.2&$\pm$8.2 & 7.46&$\pm$0.71 & 69.5&$\pm$2.8 & 5.88
&$\pm$0.29 & 76.4&$\pm$1.1 \\ 
 & K$_S$ & 5.35&$\pm$2.07 & -56.6&$\pm$33.3 & 2.22&$\pm$1.09 & 82.4&$\pm$9.8
 & 4.09&$\pm$0.34 & -44.2&$\pm$35.1 \\ 
167-317 & J & 23.94&$\pm$1.38 & 30.8&$\pm$1.7 & 11.70&$\pm$1.25 & 38.5&$\pm$1.0 & 
15.28&$\pm$2.48 & 11.9&$\pm$2.3 \\ 
 & H & 5.86&$\pm$1.48 & -10.8&$\pm$1.5 & 7.84&$\pm$2.30 & 56.8&$\pm$3.0 & 
13.34&$\pm$6.57 & 21.4&$\pm$6.4 \\ 
 & K$_S$ & 1.14&$\pm$0.75 & 85.6&$\pm$11.5 & 5.09&$\pm$0.45 & 68.4&$\pm$2.9
 & 8.26&$\pm$6.54 & 19.1&$\pm$9.6 \\ 
177-341 & J & 5.54&$\pm$1.72 & -17.0&$\pm$4.5 & 9.68&$\pm$1.64 & -6.1&$\pm$3.8 & 
20.38&$\pm$2.59 & -3.6&$\pm$0.9 \\ 
 & H & 1.64&$\pm$0.96 & 36.5&$\pm$4.1 & 10.27&$\pm$0.80 & 27.3&$\pm$1.7 & 
19.90&$\pm$2.02 & 17.4&$\pm$0.5 \\ 
 & K$_S$ & 3.42&$\pm$0.20 & 30.3&$\pm$0.3 & 0.80&$\pm$1.06 & -46.3&$\pm$33.0
 & 11.14&$\pm$0.70 & -58.0&$\pm$0.4 \\ 
178-342 & J & 17.12&$\pm$3.77 & 50.7&$\pm$0.6 & 20.48&$\pm$2.03 & 15.4&$\pm$1.6 & 
25.18&$\pm$2.87 & 7.5&$\pm$0.2 \\ 
 & H & 6.30&$\pm$2.07 & -41.3&$\pm$35.5 & 11.02&$\pm$3.93 & 16.2&$\pm$4.8 & 
23.94&$\pm$1.99 & 6.4&$\pm$0.6 \\ 
 & K$_S$ & 11.11&$\pm$3.19 & 63.5&$\pm$4.4 & 0.36&$\pm$1.14 & 87.5&$\pm$30.6
 & 3.14&$\pm$1.08 & -48.2&$\pm$34.2 \\ 
178-343 & J & 11.08&$\pm$0.91 & 59.2&$\pm$4.3 & 17.64&$\pm$2.59 & 25.8&$\pm$1.6 & 
23.57&$\pm$1.01 & 15.4&$\pm$0.5 \\ 
 & H & 11.94&$\pm$3.99 & -46.3&$\pm$0.4 & 11.94&$\pm$3.03 & 23.5&$\pm$4.2 & 
24.00&$\pm$3.99 & 8.3&$\pm$0.6 \\ 
 & K$_S$ & 10.14&$\pm$3.25 & 87.2&$\pm$0.5 & 2.35&$\pm$0.73 & -39.6&$\pm$25.6
 & 8.27&$\pm$1.35 & -62.6&$\pm$0.6 \\ 
 \hline
  \end{tabular}
 \end{table*}

\begin{figure*}
\centering
\includegraphics[height=\textwidth,angle=90]{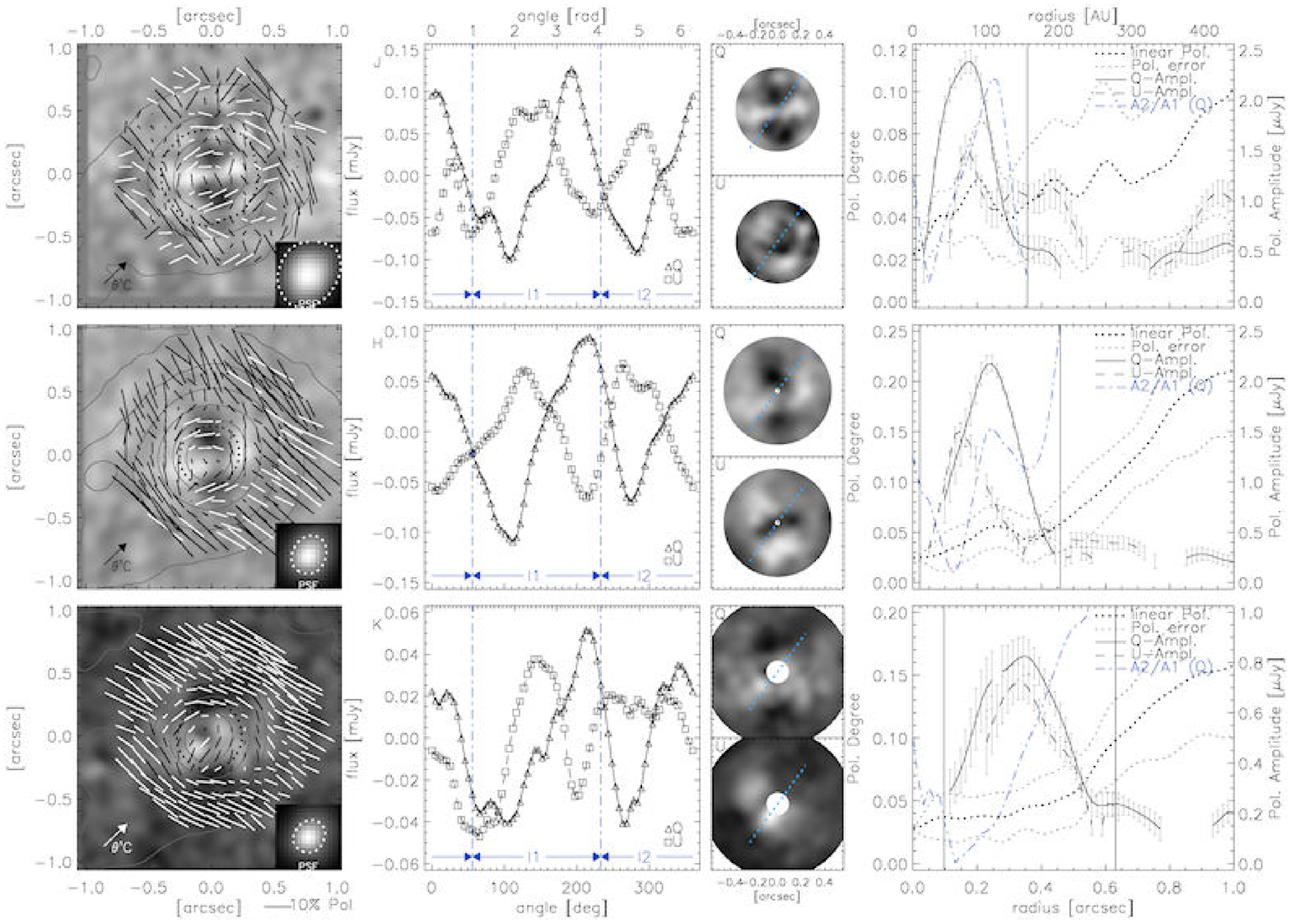}
\caption{The polarization vector maps of 177-341 show a significant pattern. Left: Stokes-Q-images with intensity 
contours (logarithmic, dotted line is 50\%) in J-, H-, and K-bands (from top to bottom), 
the vector length is proportional to the degree of polarization.
The characteristic butterfly pattern refers to an extended polarized source around the central star. 
The reflection nebula shows a uniformly oriented polarization vector pattern, which is almost perpendicular to 
the direction of $\theta^1$C Ori (indicated by the arrow in the lower left corner).
For comparison, the PSF (with dotted 50\% contour line) is plotted on the same scale in the lower right corner.
Right: Radial polarization profile, the linear polarization degree increases with the distance to the central source. 
The amplitude of the cosine-fit of the Q- and U-profiles (along the P.A.) over the radius is also plotted, 
which clearly defines the extent of the polarization pattern. The ranges are plotted with vertical lines.
Middle: Stokes-Q-profile against the angle in Jansky: 5 degree bins in the Q-component against the rotation angle in 
the range determined with the radial plot to the right and shown in the two Q/U-doughnut-plots (0$^\circ$-angle facing to the west, 
rotation counterclockwise). A slightly greater amplitude in the range of about 50$^\circ$-230$^\circ$ (interval I1) 
of the Q-profile is visible in all bands. The ratio of the fitted amplitudes in I1 and I2 (A1, A2) are also 
plotted as a function of the radius to the right 
in linear range from $0$ to $2$. 
}
\label{res}
\end{figure*}

In Fig. \ref{res} the polarization vectors patterns of 177-341 are shown. Close to the central source in the dust
envelope, a centrosymmetric
pattern is visible in relatively low polarization in the range of 5-10\%. 
The reflection nebula in the outer regions of the proplyd shows a uniformly oriented polarization pattern with the same direction in all bands.
This is almost perpendicular to the direction of $\theta^1$ Ori C, which is most likely the source of illumination of this part of the nebula.


\section{Model and discussion}
\label{model}
\subsection{Polarization properties}
\label{polprop}
Circumstellar accretion disks mark optically thick regions of the reflection nebula where the 
polarization patterns clearly depart from centrosymmetry. The vectors in these regions usually show low amplitudes of polarization
of about 10\% due to multiple scattering \citep{1988ApJ...326..334B}.
In the aperture of 0.35'' for J-band to 0.5'' for K-band, the centrosymmetric pattern is visible to the left in Fig. \ref{res}. 
A region of shortened and aligned vectors is not clearly resolvable in this view.

The outer regions show higher polarization and an aligned pattern from the illuminated reflection nebula from $\theta^1$ Ori C.
The extent of the polarization pattern around 177-341 is determined in the radial plots to the right of Fig. \ref{res} as follows.
The fraction of polarized light as a function of the distance from the star shows a plateau of low polarization
of the circumstellar envelope and is increasing for higher radii due to the nebula polarization. 
For a perfect centrosymmetry, the profiles along the position angle of the Q- and U-images should resemble a cosine.
The amplitude of the cosine fitted to the Q- and U-profiles as a function of the distance then clearly defines the extent.
The maximum radius increases from J- to K-band and a small plateau around the maximum extent is visible in all bands.  

For a closer look at the spatial proportions, we examined the profile of the Q- and U-images 
by adding 5-degree bins over the determined radial range. 
These azimuthal Q-profiles show the characteristic cosine or W-curve for a dust envelope (Fig. \ref{res}, middle).
In our data the profiles are not symmetric around zero as they should be, and this is 
equivalent to an offset in polarized flux and is related to effects of the PSFs.
The long distance of about 26'' from 177-341 to the AO guide star $\theta^1$ Ori C (Table \ref{tabprop})
results in elongated PSFs for the observations \citep{2005A&A...438..757C}. 
The direction of elongation 
is towards $\theta^1$ Ori C (134$^\circ$ P.A.).
A shifting of the profiles can be seen, after convolving simulated polarization images with the PSF of the observations 
(Figs. \ref{shiftvectors} and \ref{elongeffects}, only available electronically);
therefore, the profiles of the convolved images of the data 
and simulations were shifted to be symmetric around zero so as to achieve a result closer to the
actual centrosymmetric polarization pattern. This can be seen as a bootstrap calibration of 
the polarimetric images on the basis of the centrosymmetry of the envelope polarization.
This has an impact on the pattern of the surrounding reflection nebula, which is highly polarized, but the
total flux is low. The relative changes in the region of centrosymmetry are irrelevant due to the higher flux, while the analysis of the 
individual Q- and U-images in this study is not affected.

Since the Q- and U-images are not acquired simultaneously, there are deviations due to residual
speckles and changes in seeing in addition to the elongated PSFs. 
These errors are emphasized during the determination of $\Pi_L=\sqrt{Q^2+U^2}$.
As a consequence, no clear features are seen in the polarization degree images, and no direct comparison to
convolved simulation images is possible. Hence, we tried to deconvolve the observational data in this case.
For the PSF estimation the FOV with 177-341 was analyzed with the StarFinder program \citep{2000A&AS..147..335D}.
Three point-like sources around 177-341 were used to extract a field PSF
(Fig. \ref{psfestim}, only available electronically). For the deconvolution of the individual intensity images $I_\phi$,
the iterative Lucy-Richardson
\citep{1974AJ.....79..745L}
algorithm was chosen, which is known to be robust and stable even under difficult SNR conditions 
\citep{1998ASPC..145..496P}.
For the beam reconstruction, the deconvolved $I_\phi$-images were convolved with a Gaussian of 0.25'' FWHM, before
the polarization degree images were generated.
The diffraction limit at K-band is 0.15'' for the 3.6m telescope, but under 
moderate seeing conditions and these low Strehl ratios of the observations (Sect. \ref{chap:obs}),
an empirical limit for the achieved resolution is 1.5--2 diffraction limit units \citep{1997SPIE.3126..589B}.
Without this smoothing, the deconvolved images are also dominated by high-frequency noise features 
enhanced progressively with the number of iterations \citep{1999A&AS..140..235L}.
The algorithm converges fastest at the brightest source structures \citep{1974AJ.....79..745L}.
Therefore choosing too few iterations results in images of undefined resolution and quality.
From our extensive experience with the algorithm \citep{1999ApJ...523..248O}, we chose $10^3$
iterations for the case of a single star dominating a single circumstellar structure.

\begin{figure*}
\centering
\includegraphics[height=\textwidth,angle=90]{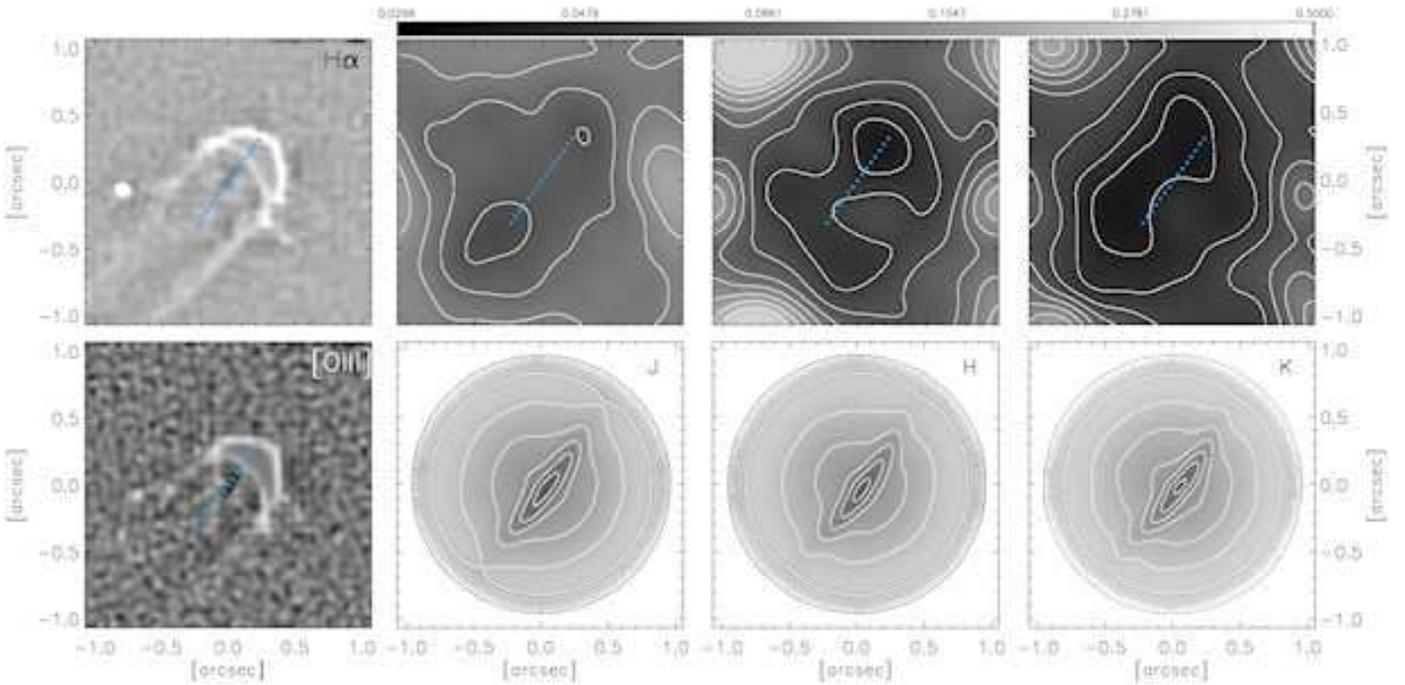}
\caption{Left: HST optical images \citep{2000AJ....119.2919B} of 
177-341 through the filters f656N (\ion{H}{$\alpha$}, top) and f502N (\lbrack\ion{O}{III}\rbrack, bottom).
The images have been subtracted by a Gaussian-smoothed version of itself, which enhances the local contrast and attenuates 
extended emission.
A possible disk orientation is marked. Right top: Polarization degree images of 177-341 in J-, H-, and K-bands (from left to right).
The individual intensity images $I_\phi$ were deconvolved before the calculation of the polarization degree, which
is very sensitive to PSF-effects.
Right bottom: 
Polarization degree images for a MC simulation with disk inclination of 75$^\circ$. 
}
\label{pa}
\end{figure*}

For the position angle of the disk, we compared the HST optical images and simulations to 
the polarization degree maps, obtained from deconvolved intensity images (Fig. \ref{pa}). The parameters of the simulation 
are discussed in Sect. \ref{simchap}, but here a model with highly polarized envelope is shown to
illustrate the disk signature in the envelope signal.
In the optical images a small portion of the disk is seen in silhouette around the central star. Almost the same
P.A. is found as an elongated region of low polarization in the NIR polarization images and can be  
quantified to $\sim144^\circ$. The diameter of the disk is about 300AU in this view. 
It must be stated that the direction of elongation of the PSFs differs only $10^\circ$ from the obtained P.A. and
the elongated low polarization region could just be an artefact. But the extent of the region of 300AU clearly exceeds the
PSF dimensions at least in the K-band. The P.A. is also confirmed by the following analysis.

\begin{figure*}
\centering
 \begin{tabular}{c}
\includegraphics[width=0.48\textwidth]{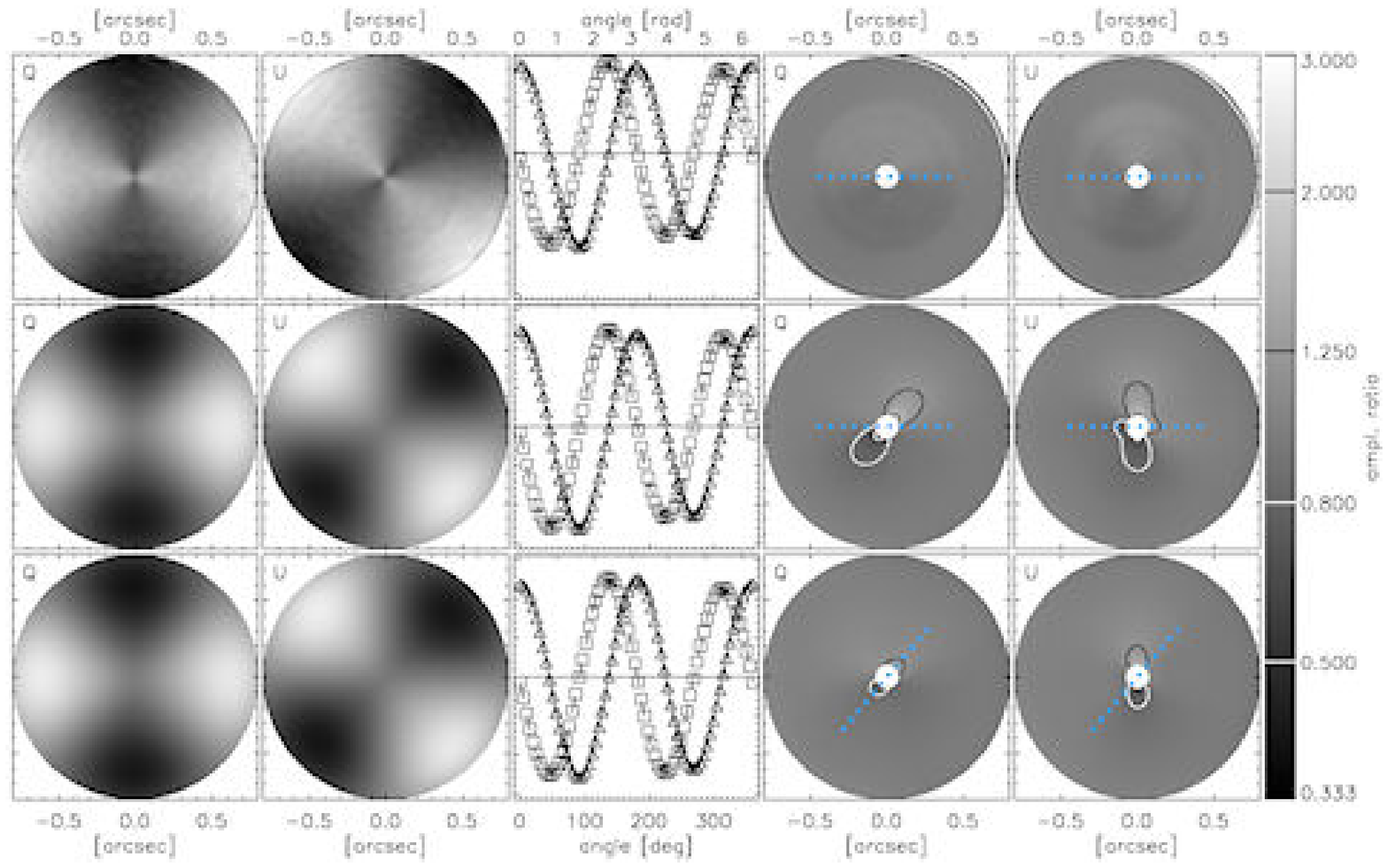}
\includegraphics[width=0.48\textwidth]{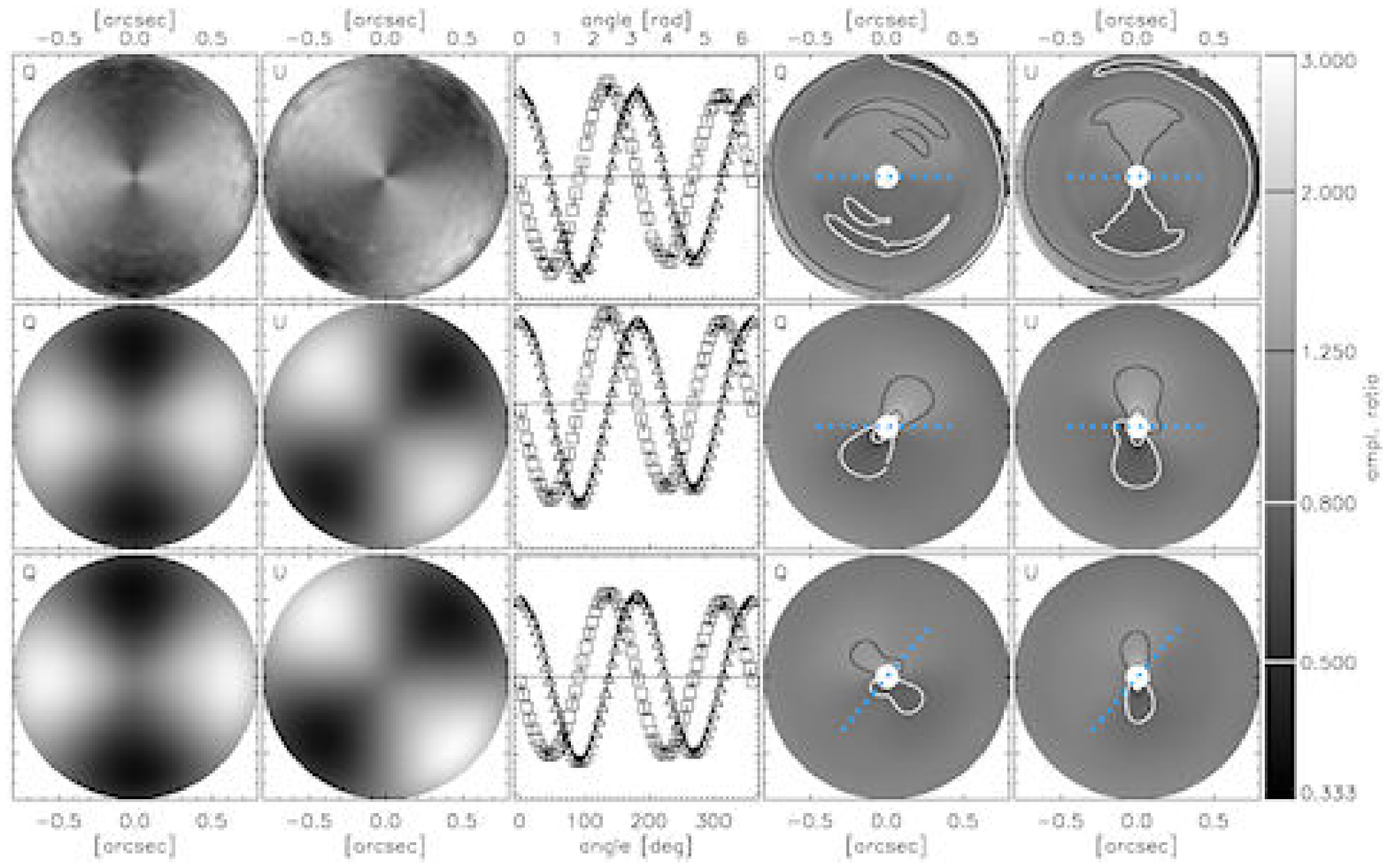}
\end{tabular}
\caption{Monte Carlo simulations of circumstellar envelope and disk. Left panel from left to right: Q-image, U-image, Q- and U- profile,
Q-amplitude ratio map, and U-amplitude ratio map in H-band. The disk is dense and flat ($M=0.1M_{\odot}$, $h=0.001R_S$). The inclination is 
$\xi=10^{\circ}$. From top to bottom: unconvolved images with P.A.$=0^{\circ}$ for the disk, images convolved with elongated PSF,
P.A.$=0^{\circ}$ and convolved images with P.A.$=50^{\circ}$. 
Right panel: the same for a dense and flared disk ($M=0.1M_{\odot}$, $h=0.01R_S$).
The profiles and maps show no significant modulation
in the nearly edge-on configuration. }
\label{ratioexplan1}
\end{figure*}

\begin{figure*}
\centering
 \begin{tabular}{c}
\includegraphics[width=0.48\textwidth]{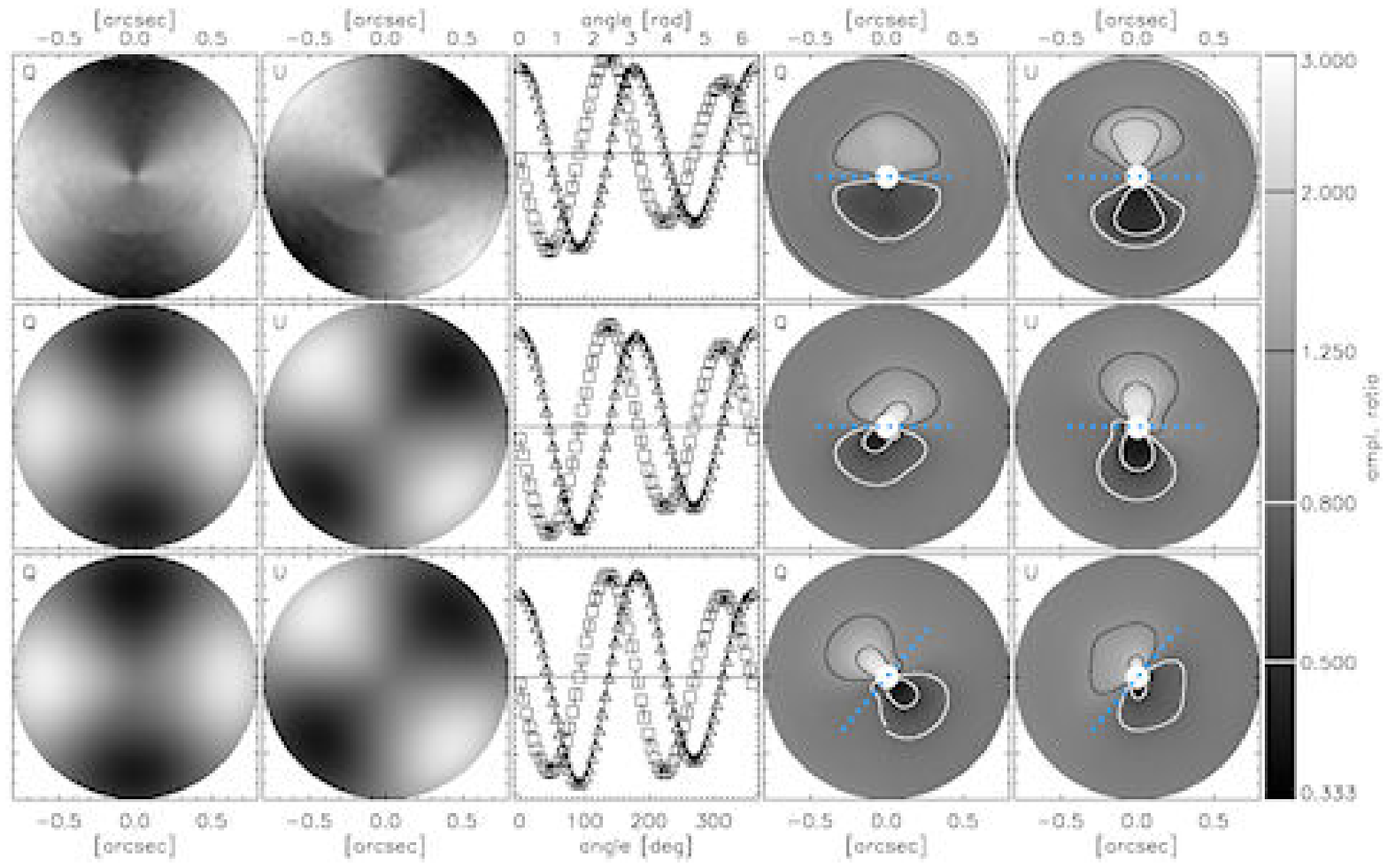}
\includegraphics[width=0.48\textwidth]{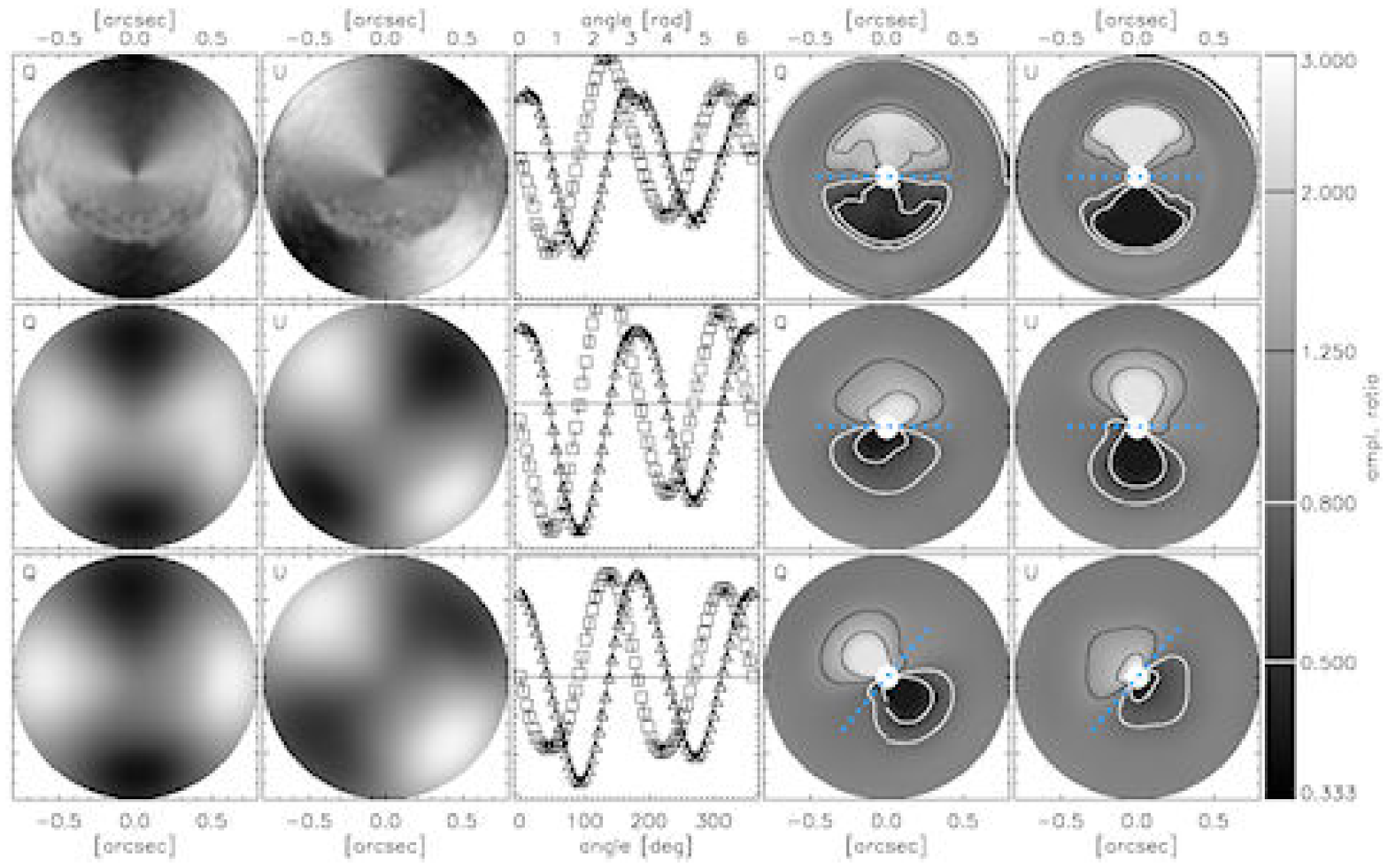}
\end{tabular}
\caption{Same as Fig. \ref{ratioexplan1}, but for $\xi=45^{\circ}$. 
The effects of the disk on the envelope signal is strong, as the amplitude of the profile is attenuated
where the disk lies in front. In the maps, the P.A. is clearly seen and the radial extent of the modulation
refers to the inclination angle.}
\label{ratioexplan2}
\end{figure*}

\begin{figure*}
\centering
 \begin{tabular}{c}
\includegraphics[width=0.48\textwidth]{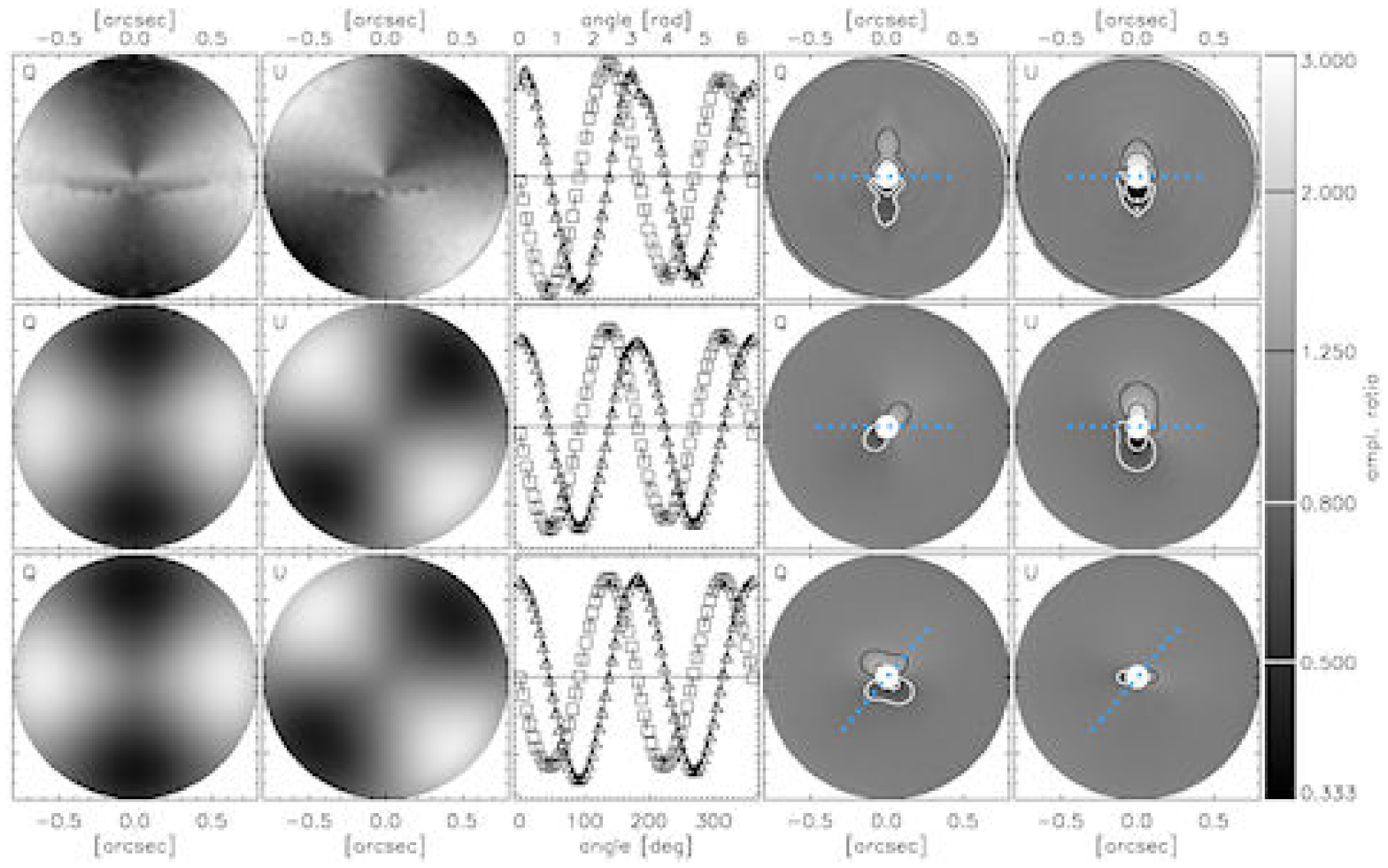}
\includegraphics[width=0.48\textwidth]{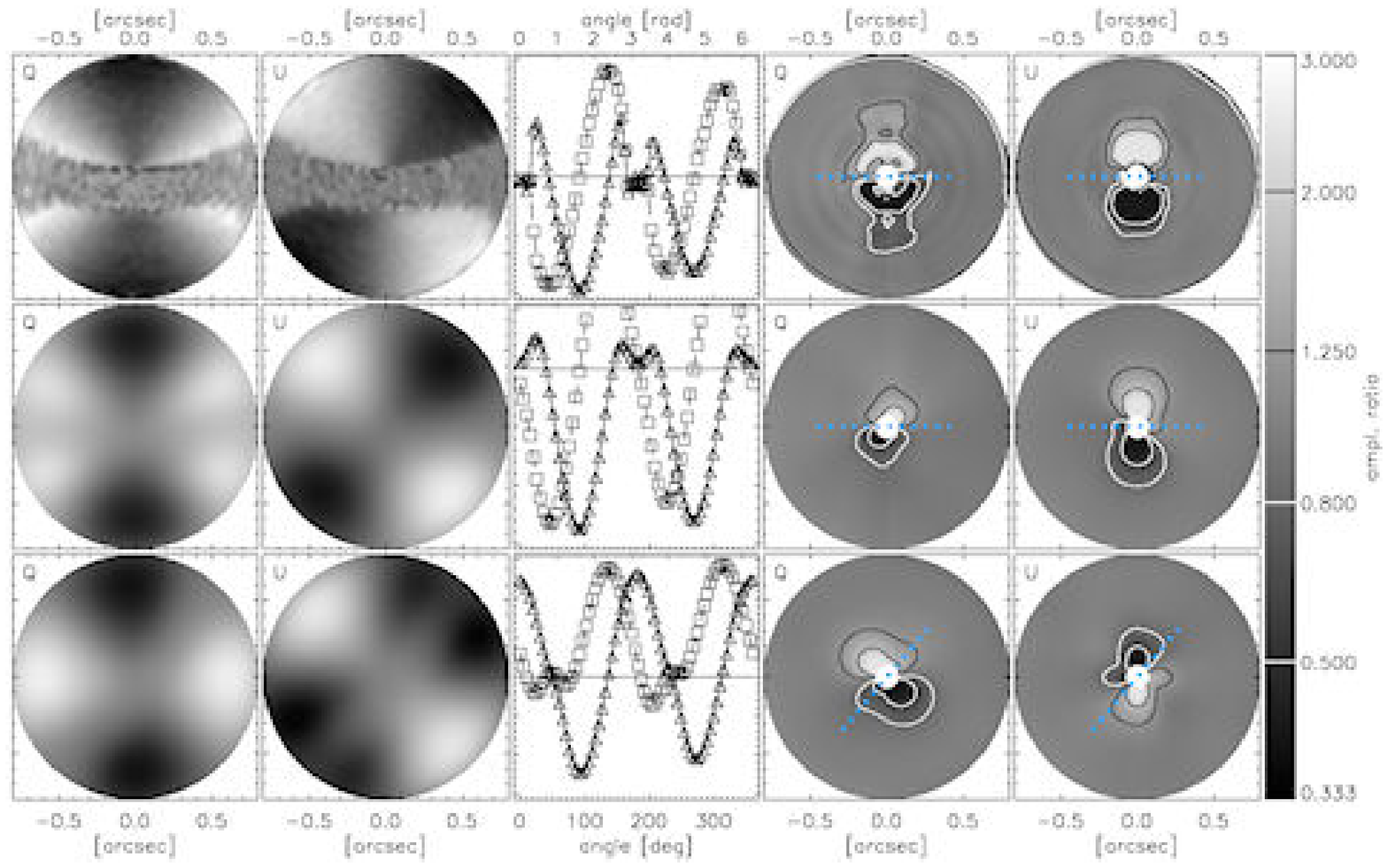}
\end{tabular}
\caption{Same as Fig. \ref{ratioexplan1}, but for $\xi=80^{\circ}$. Nearly seen edge-on, the disk modulation
is still seen in the envelope signal. For the flat disk (left panel) the extent of the modulation refers
to the inclination, as the height of the disk is negligible against the minor axis of the ellipse in projection.
The extent of the highly flared disk (right panel) is dominated by the disk height.}
\label{ratioexplan3}
\end{figure*}

There is a noticeable modulation of the amplitudes of the profiles, which can constrain disk parameters if we consider the optically
thick accretion disks. 
In regions where the disk lies in front of the illuminating source from the observer's point of view the amplitude should be attenuated. The same
applies where the disk blocks the propagation of the central light to outer regions of the envelope.
This can be associated with the amplitude modulation (AM) technique used for communication and signal transmission.
The envelope scattering is the carrier signal, i. e. a cosine with fixed phase in the Q- and U-profiles. 
Since there is no reference amplitude and we have only two cycle periods (see Fig. \ref{res}), only the ratio between the
two can be determined as the modulation signal.
To investigate these effects, we generated amplitude ratio maps, which are defined in polar coordinates as follows.
For every position angle and radius ($r_0$, $\phi_0$) the Q-profile over a
range from ($r$ ,$\phi_0-90^\circ$) to ($r$, $\phi_0+90^\circ$) is fitted with a function $f(\phi)=A_0+A\cos{(2\phi-\phi_Q)}$ 
with free amplitude $A$ and amplitude offset $A_0$,
but fixed phase offset $\phi_Q=0^\circ$ and $\phi_U=45^\circ$, respectively.
The best-fitting amplitude is divided by the one of the remaining $180^\circ$ 
(for clarification see Fig. \ref{res}, middle: dash-dotted lines).
The cosine-fitting incorporates the model of the fixed carrier signal amplitude. The individual fitting over the radius $r$
allows arbitrary radial density and polarization profiles for the envelope, but rotational asymmetries like flattened envelopes are not considered. 
The result has an
advantage over the pure Q-/U-profiles, since the attenuation of the signal is determined over the whole position angle,
and the SNR is increased under the assumption of centrosymmetric envelope profiles.
In principle the achievable spatial resolution is also enhanced by the high sensitivity of the signal to modulations.
The results are shown in Figs. \ref{ratioexplan1}--\ref{ratioexplan3} for models
with a flat and a flared disk, respectively. The behavior of the maps for the unconvolved images is as expected and
can separate the disk signal from the envelope signal to some extent.
A ratio map of a flat disk indicates the inclination angle. In this case the disk silhouette forms an ellipse
with semi-major axis $a$ (the disk radius) and semi-minor axis $b$, so the inclination is given by $\xi=\arccos{(b/a)}$.
Because the radial extent of the modulation in the amplitude ratio map  
refers to $b$, the inclination can be determined.
However, if the images are convolved with elongated PSFs, the Q- and U-maps perform differently, depending on the P.A. of the disk.
For a P.A. of $0^\circ$, the Q-map is distorted, because of the low amplitude of the carrier in the region of the modulation by the disk
(Fig. \ref{ratioexplan3}, left panel, e.g.). The same applies for a P.A. of $\sim 45^\circ$ and
the U-map. With a low-amplitude signal, PSF elongation effects become dominant.

\begin{figure}
\centering
\includegraphics[height=0.5\textwidth,angle=90]{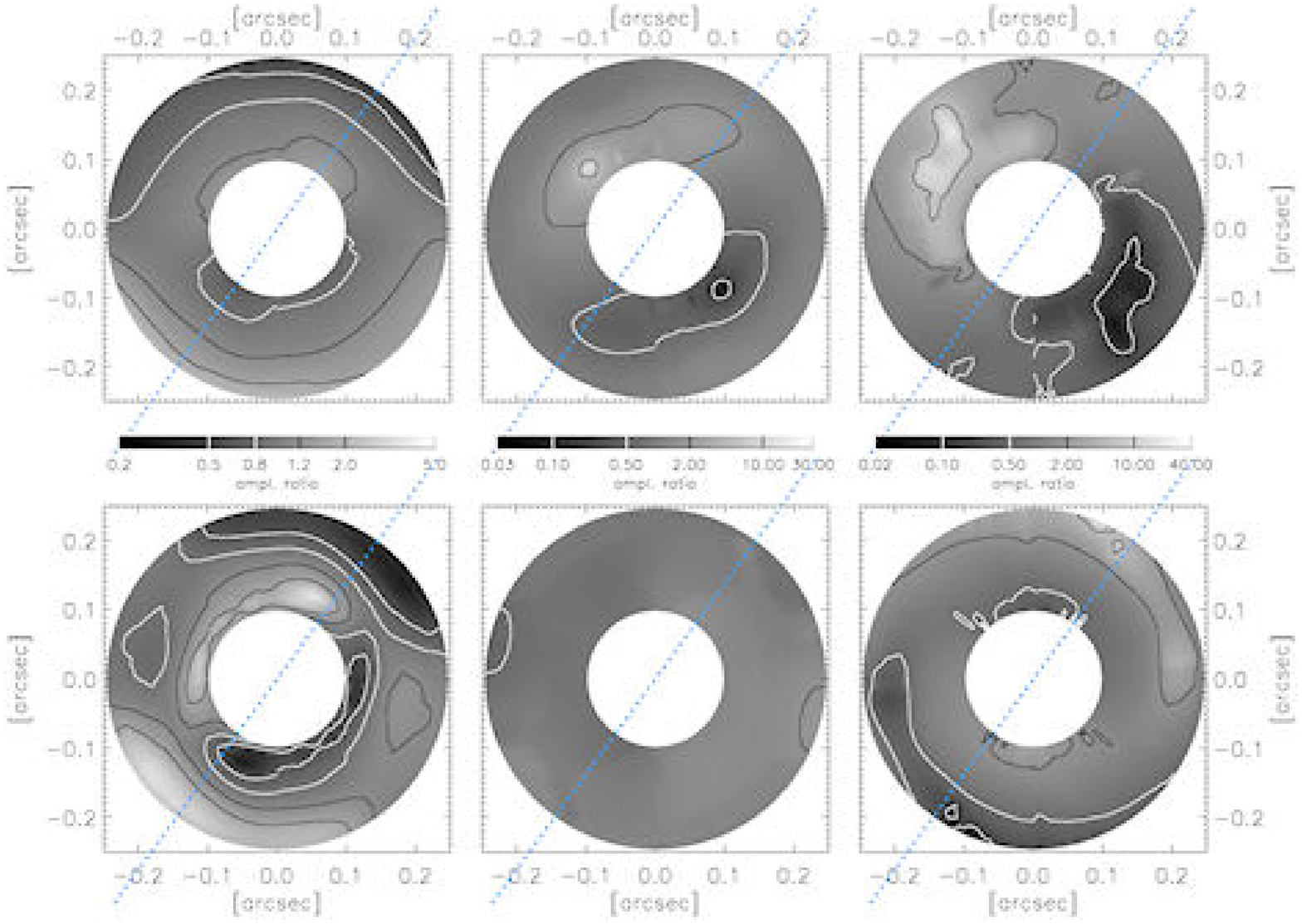}
\caption{Plot of the test of the inclination of the disk: the amplitudes 
of the polarization profile (see Fig. \ref{res}) are attenuated due to the multiple scattering in the disk.
If the disk is inclined, the amplitude of the part with the disk in front is more affected. 
Therefore the ratio of the amplitudes (``in front'' divided by ``behind'' ) is minimal. 
Here the maps of the Q- (top) and U-amplitudes ratio in J-, H-, and K-bands (from left to right) are shown in an annulus of 0.1''-0.25''. 
The position angle for the disk can be located around 144$^\circ$. 
The radius of the minimum spots ratio along the line perpendicular to the disk can be determined to $\sim$0.12'' in Q (H- and K-bands).
}
\label{posan}
\end{figure}

The maps of the amplitude ratio of the observations in J-, H-, and K-bands are shown in Fig. \ref{posan}.  
In the Q-maps we find extremal ratios in the H- and K-band at a position angle of $144^\circ$ of the disk, so the Q-map should be less
affected by elongation effects. The modulation in the U-maps is not significant -- note the individual scaling
of each band -- and is interpreted as PSF- and noise effects. This is clear evidence that the disk
is inclined and the portion in front of the star is located in the south west. Moreover, the radius $b$ of the minima can be determined 
precisely to 0.12'' in the H-band and in K-band 
(Figs. \ref{posan} and \ref{res} right, dash-dotted line). 
If a radius of $a=0.35''$ and a flat disk are assumed, 
the shape of the ellipse is set and the minimum inclination angle
is given by $\xi=\arccos{(b/a)}\approx70^\circ$. Considering the height of the (flared) disk, this
value is increased. This can be seen in Fig. \ref{ratioexplan3}, where the flared disk produces a greater
extent as the flat disk for the same inclination.
This results in an estimated possible inclination range of about 75$^\circ$--85$^\circ$, as the disk
has to be inclined according to the signature in the ratio maps.
The values of maximum ratios increase from the J- to the K-band. 
This can be interpreted as a stronger relative attenuation of the disk at
higher wavelength. Since the polarization is lower in K-band, the total attenuation might be constant.

\subsection{Modeling}
\label{simchap}
There are two main processes for producing NIR/optical polarization: (i) scattering and (ii) 
differential absorption by non-spherical dust grains with the short axis
primarily aligned along the local magnetic field, referred to as
dichroic extinction \citep{1988ApJ...326..334B}.
In the case of a circumstellar envelope and disk, the first is relevant. 
To interpret the polarization data, we used a 3D radiation transfer code, 
based on the Monte Carlo method \citep{1992ApJ...395..529W,1993ApJ...402..605W}. 
The model parameters of the presented simulations, which 
illustrate the signature of the disk in the polarization images
(Figs. \ref{pa}--\ref{posan}), are described in the following. In addition
different models for grain size distributions and the impacts on the polarization images are discussed
as an attempt to explain the increasing polarization pattern from the J- to the K-band.

The standard models for low-mass star formation include a slowly rotating, isothermal dense molecular cloud that collapses.
The material with high angular momentum forms the circumstellar disk, while the one with lower angular momentum falls into the proto star.
We adopted the model of an infalling proto-stellar cloud described by \cite{1984ApJ...286..529T}, often referred to as the TSC-model.
Our configuration includes both a circumstellar envelope and a disk. We skipped a 
detailed modeling of bipolar cavities, since we do not see any clear evidence in the data. 
As mentioned in \cite{1993ApJ...402..605W}, an infalling envelope can easily produce the specific 
polarization pattern of reflection nebulae, which are 
often interpreted as scattering from a disk. As a consequence we expect a highly inclined disk in the case of 177-341 as suggested by
the optical images (Fig. \ref{pa}), although we see a round, nearly centrosymmetric pattern in the polarimetric images. 
The critical parameter for reproducing the pattern is the mass infall rate in the TSC-model. 
As the envelope size is
limited to about 300AU in our images (Fig. \ref{res}), the centrifugal radius is determined to 150AU in the model.

\begin{table*} 
  \caption{Dust properties of the standard models MRN and KMH. }
\label{graintab}
 \centering
  \begin{tabular}{lccccccccccccccc}
  \hline\hline
 && \multicolumn{4}{c}{MRN}&&\multicolumn{4}{c}{KMH}&&\multicolumn{4}{c}{KMH$_{DC}$} \\
\cline{3-6} \cline{8-11} \cline{13-16} 
Band& &$\kappa$ & $\omega$ & $g$ & $P_{\mbox{max}}$& &$\kappa$ & $\omega$ & $g$ & $P_{\mbox{max}}$& &$\kappa$ & $\omega$ & $g$ & $P_{\mbox{max}}$\\
  \hline
J&&65&0.42 &0.16 & 0.81 &  & 63 & 0.46 & 0.32 & 0.58 &  &  48  & 0.60 & 0.34 & 0.58 \\  
H&&38&0.33&0.06&0.91 &     &  38 & 0.42& 0.29 & 0.59 &  &   33  & 0.56 & 0.31 & 0.59 \\    
K&&20&0.21&0.03&0.94 &     &  22 & 0.36 & 0.25 & 0.60 &  &  20 & 0.50 & 0.27 & 0.60 \\      
\hline
   \end{tabular}
\begin{list}{}{}
\item[Note:] The unit of $\kappa$ is [cm$^2$g$^{-1}$].
\end{list} 
 \end{table*}

The dependency of polarization intensity on the wavelength refers to the effective grain size in the dust envelope, which
scatter and polarize the emission from the central star \citep{1978A&A....70L...3E}. 
The polarization spectrum and peak polarization also depend strongly on the albedo $\omega$, the opacity $\kappa$,
and the asymmetry parameter $g$ of the grains. The latter defines the relation between forward scattering and isotropic scattering.
For the grain size distribution and parameters, we adopted the models from \cite{1977ApJ...217..425M}, hereafter MRN and 
\cite{1994ApJ...422..164K}, KMH. 
Table \ref{graintab} gives an overview of the differences between the two 
commonly used models in the NIR (data from \cite{1994ApJ...422..164K} and \cite{1997ApJ...485..703W}). 
In MRN, grain sizes are limited to $0.25\mu m$, which is
small compared to the wavelength of NIR-light, so the albedo and opacity drop
for increasing wavelength. The KMH-model uses grain parameters obtained from fits to the interstellar extinction law.
Local variations in the interstellar extinction curve are parameterized with the ratio of total to selective extinction $R_V\equiv A(V)/E(B-V)$.
While the standard KMH deals with the diffuse interstellar medium ($R_V=3.1$), the alternative (here denoted as KMH$_{DC}$) is calculated
for a dense cloud region with $R_V=5.3$. 
\cite{1981ApJ...244..483M} stated that (i) an increasing $R_V$ implies an increasing size of particles, and (ii) $\lambda_{\mbox{max}}$,
the wavelength of maximum polarization, is larger for regions with $R_V$ larger than the average $3.1$. The peculiar extinction
of $\theta^1$ Ori C is specified with $R_V=5.5$. 
From the fit to the modified extinction curve in KMH$_{DC}$, the $R_V=5.3$ distribution contains significantly
fewer small- and intermediate-sized particles and a modest increase in larger sizes \citep{1994ApJ...422..164K}. 
This implies both a higher albedo and slightly greater $g$ and therefore an
increased forward scattering. The extinction is less at short wavelengths and flattened towards higher wavelengths. 
We also tested a modified model with different grain properties only for the disks region.
If grain growth in the disk is achieved, the scattering properties here should differ from the envelope.
We adopted calculations from \cite{1990ApJ...349..107P} for large grains with a $\sim$4 times higher opacity and higher albedo,
but also implemented a wavelength dependency, so that the scattering in K-band is increased.
With high opacities and thus a high degree of multiple scattering, photons could be reprocessed by the disk more easily. 
Therefore more photons would be available in the outer regions of the envelope, so the scattering
toward the observer could be enhanced. 

The parameters for the envelope and the disk are chosen as determined from the data in the previous Section
and listed in Table \ref{tabmc}.
This includes the envelope and disk size, and the inclination and P.A. of the disk. The mass infall rate for
the envelope was probed in a wide range of $10^{-9}$--$10^{-6}$ $M_\odot y^{-1}$ to vary the visible extent
of the polarization pattern. A value of about $10^{-9}$ gives the best results for the average radius of 0.48''.
Such low infall rates would refer to a low-mass proto star in a transition between late Class I (embedded in envelopes)
and Class II (with T Tauri disks) \citep{2003ApJ...598.1079W}.

 \begin{table} 
 \caption{Parameters used in Monte Carlo simulation.} 
 \label{tabmc}
 \centering
  \begin{tabular}{lc}
  \hline\hline
Parameter & Value\\ \hline
 Number of photons& 10.000.000\\
 Stellar radius $R_S$ & 1.5$R_\odot$\\
 Disk mass $M$ & 0.1$M_\odot$\\
Outer disk radius & 150AU\\
 Inner disk radius & 4$R_S$\\
 Inclination angle & $75^\circ$\\
 Position angle & $144^\circ$\\
 Scale height of disk $h$& 0.001$R_S$\\
 Inner envelope radius & 10$R_S$\\
 Outer envelope radius & 150AU\\
 Mass infall rate for envelope& $1\cdot 10^{-9} M_\odot y^{-1}$\\ 
  \hline
   \end{tabular}
\begin{list}{}{}
\item[Note:] The spatial dimensions are derived from the observational data as described in Sect. \ref{polprop}.
\end{list}
 \end{table}

The simulation images are convolved with the PSFs from the observational data to compare the results.
The convolution with the elongated PSFs causes a change in the 
polarization properties: an offset of polarization of the position angle in the direction of the elongation is now visible
(Fig. \ref{shiftvectors}, only available electronically). 
This concurs with a shift in the Q- and U-profiles as seen in the centrosymmetric regions of the data
(Fig. \ref{res} and Sect. \ref{polprop}).
The amount of offset depends on the specific
model parameters, particularly the densities of disk and envelope and the disk inclination.

\begin{figure}
\centering
\includegraphics[height=0.45\textwidth,angle=90]{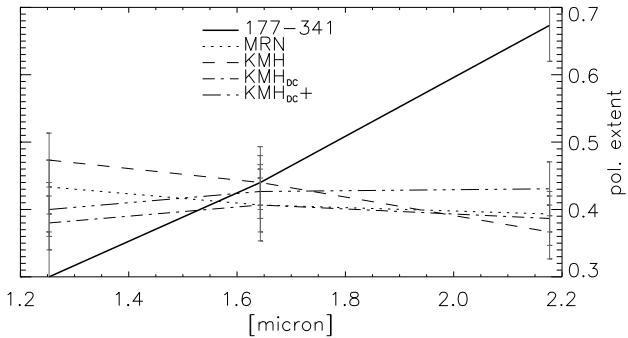}
\caption{Results from simulations with different grain models in comparison to the observational data.
The polarization extent in arcsec as a function of wavelength is shown (solid line, determined as in Fig. \ref{res}).
The standard models MRN (dotted line) and KMH (dashed line) show a reversed spectral dependency. KMH$_{DC}$ 
(dash-dotted lines) for a dense cloud is flattened.
}
\label{simres}
\end{figure}

\subsection{Discussion}

\cite{1999AJ....118.2350H} derived an inclination angle of the disk of 177-341 of 75$^\circ$--85$^\circ$
by fitting photo-evaporation model data to emission line profiles of spectroscopic data.
\cite{2002ApJ...578..897R} modeled the SED for a photo-ionized proplyd and found that the
disk emission is bright in the infrared and comparable to the central star.
They also propose a highly inclined disk
configuration for 177-341 by comparison to 10$\mu$m-images. \cite{2000AJ....119.2919B} interpreted HST 
\lbrack\ion{O}{III}\rbrack-images of 177-341 as a
disk seen in silhouette against the bright IF. They propose a P.A. of $105^\circ$, although the silhouette suggests
a P.A. of about $135^\circ$ to $145^\circ$ because of the compact source about 0.5'' east of 
this proplyd, which is likely to be an ultra compact Herbig-Haro object (177-341b in Fig. \ref{ifoptn}). 
\cite{2005A&A...441..195V} performed a deep analysis on the HST data and determined the disk size from the IF diameter. 
For a distance of 0.01--0.3 pc to $\theta^1$ Ori C the FUV (far-ultraviolet) photons dominate the dissociation process 
and the disk radius is typically $r_d \gtrsim r_{IF}$ \citep{1998ApJ...499..758J,1999ApJ...515..669S}. In the case of 177-341
the diameter of the IF is $\sim$0.8" or 350 AU.

Considering these constraints on the disk and envelope configuration we find good agreement of the parameters 
with our polarimetric analysis. 
The compliant diameter and P.A. justify the deconvolution process and the introduced polarization
amplitudes ratio maps. The inclination range obtained is also consistent with
previous estimates. In the special case of a polarized envelope and a circumstellar disk, 
polarimetric differential images can be interpreted with this technique even under poor SNR and AO conditions. 
With the obtained values for the envelope size and the size, P.A. and inclination of the disk the simulations 
show similar polarization images.

According to a common model of the eroding disk mechanism 
\citep{2003ASPC..287..263B}, first the outer parts of the disk are being destroyed. One way
for disks to survive the photo-evaporation by UV-radiation is for the dust grains to grow to sizes with radii $\gtrsim1$cm.
Evidence of grain growth in circumstellar disks in Orion has previously been found, because
the outer portions of the giant disk of 114-426 contains grains $\gtrsim2\mu$m \citep{2003ApJ...587L.109S}. 
The efficiency of grain growth is predicted to be highest in the center of the disk where the highest
densities and temperatures are found and
photo-evaporation does not operate efficiently \citep{2001Sci...292.1686T}.

However our comparison of different dust models cannot explain the peculiar spectral polarization of 177-341.
The models for dense clouds with high $R_V$ result in improved scattering at longer wavelengths,
but do not reproduce the increasing polarization pattern extent from the J- to the K-band (Fig. \ref{simres}). 
Also a simple model (KMH$_{DC}+$) of large grains in the disk shows only a moderate increase in the extent.
As a conclusion, advanced modeling is required including significantly different dust properties.
In particular the opacity distribution within the envelope might be worth a detailed investigation in
future model improvements.
\cite{2008ApJ} recently showed discrepancies between the extent of dust continuum emission and molecular gas emission, suggesting
a different shape for the edge of the disk. An alternative disk model with a tapered edge 
agrees more with observations than the commonly used truncated power-law model. 

\section{Summary and conclusions}
\begin{figure}
\centering
\includegraphics[width=0.45\textwidth]{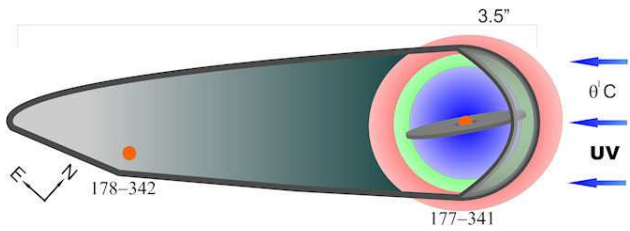}
 \caption{Schematic overview of our model and results for the proplyds 177-341 and 178-342. 177-341 has an inclined disk and an
illuminated dust envelope. 
The diameter, inclination, and P.A. of the disk and the size of the envelope are drawn according to the results of the polarimetric analysis. 
}
\label{cart}
\end{figure}

We have presented polarized emission maps of the circumstellar dust distribution of the giant proplyd 177-341 and 
found a complex morphology including both a circumstellar envelope and an accretion disk. 
The size of the disk was estimated with the diameter of the IF and 
confirmed in the polarization degree images. We extracted the inclination of the disk from a detailed study of
the polarization profiles in the Stokes images. For the P.A. of the disk, we presented evidence of part of the disk
seen in silhouette in the optical images, and we find matching position angles in the polarization amplitude ratio 
and in the deconvolved polarization-degree images due to
multiple scattering in the disk.
An overview of the results of the envelope and disk configuration is given in Fig. \ref{cart}.
The size of the dust envelope  
is obtained from the polarimetric differential images (Fig. \ref{res}) and increases from the J- (blue)
to the K-band (red). In the H-band it is comparable
to the diameter of the ionization fronts head in the optical (Fig. \ref{ifoptn}, right).
178-342 has no resolvable dust distribution, as the images in Fig. \ref{resrgb} show and modulates
the shape of the IF of 177-341 in the optical (Fig. \ref{ifoptn}).
The wavelength dependency of polarization intensity and the extent of the polarization pattern
cannot be reproduced in the models by large grains in the circumstellar material. 
The comparison of common dust models 
shows evidence of the contribution of larger grain to polarization and scattering effects, but another mechanism
must be present to explain the observational results.

We showed with our 4 m-class data that the technique of ground-based polarimetric differential imaging
is suitable for mapping the dust emission very close to the central star (here about 0.1''). 
This is the case for the largest and brightest proplyds. Other proplyds close
to the Trapezium might be worth observing with an 8 m-class telescope. 
The proplyds 158-326 and 166-316 are the most suited candidates, 
as these show extended emission and a modulated trefoil pattern in the polarimetric differential images.
With the introduced PDI analysis techniques, we are able to constrain
disk parameters, such as inclination, P.A., and size even when the disk structure itself is not resolvable.

\nocite{2001ARA&A..39...99O,
2006ApJ...644L..71S,
2006ApJ...642..339S,
2001Sci...292.1686T,
2003ApJ...587L.109S,
1994AJ....108.1375S,
1994AJ....108.1382M}

\begin{acknowledgements}
      This work was supported in part by the 
      \emph{Deut\-sche For\-schungs\-ge\-mein\-schaft (DFG)} via grant SFB 494.
\end{acknowledgements}

\bibliographystyle{aa}  
\bibliography{7555}   

\begin{thebibliography}{52}
\expandafter\ifx\csname natexlab\endcsname\relax\def\natexlab#1{#1}\fi

\bibitem[{{Ageorges} {et~al.}(1997){Ageorges}, {Eckart}, {Monin}, \&
  {Menard}}]{1997A&A...326..632A}
{Ageorges}, N., {Eckart}, A., {Monin}, J.-L., \& {Menard}, F. 1997, \aap, 326,
  632

\bibitem[{{Ageorges} \& {Walsh}(1999)}]{1999A&AS..138..163A}
{Ageorges}, N. \& {Walsh}, J.~R. 1999, \aaps, 138, 163

\bibitem[{{Bally}(2003)}]{2003ASPC..287..263B}
{Bally}, J. 2003, in ASP Conf. Ser. 287: Galactic Star Formation Across the
  Stellar Mass Spectrum, ed. J.~M. {De Buizer} \& N.~S. {van der Bliek},
  263--274

\bibitem[{{Bally} {et~al.}(2000){Bally}, {O'Dell}, \&
  {McCaughrean}}]{2000AJ....119.2919B}
{Bally}, J., {O'Dell}, C.~R., \& {McCaughrean}, M.~J. 2000, \aj, 119, 2919

\bibitem[{{Bally} {et~al.}(1998){Bally}, {Sutherland}, {Devine}, \&
  {Johnstone}}]{1998AJ....116..293B}
{Bally}, J., {Sutherland}, R.~S., {Devine}, D., \& {Johnstone}, D. 1998, \aj,
  116, 293

\bibitem[{{Bastien} \& {Menard}(1988)}]{1988ApJ...326..334B}
{Bastien}, P. \& {Menard}, F. 1988, \apj, 326, 334

\bibitem[{{Bonaccini} {et~al.}(1997){Bonaccini}, {Prieto}, {Corporon},
  {Christou}, {Le Mignan}, {Prado}, {Gredel}, \& {Hubin}}]{1997SPIE.3126..589B}
{Bonaccini}, D., {Prieto}, E., {Corporon}, P., {et~al.} 1997, in Presented at
  the Society of Photo-Optical Instrumentation Engineers (SPIE) Conference,
  Vol. 3126, Proc. SPIE Vol. 3126, Adaptive Optics and Applications, Robert K.
  Tyson and Robert Q. Fugate, Eds., p.589, ed. R.~K. {Tyson} \& R.~Q. {Fugate},
  589--+

\bibitem[{{Cresci} {et~al.}(2005){Cresci}, {Davies}, {Baker}, \&
  {Lehnert}}]{2005A&A...438..757C}
{Cresci}, G., {Davies}, R.~I., {Baker}, A.~J., \& {Lehnert}, M.~D. 2005, \aap,
  438, 757

\bibitem[{{Diolaiti} {et~al.}(2000){Diolaiti}, {Bendinelli}, {Bonaccini},
  {Close}, {Currie}, \& {Parmeggiani}}]{2000A&AS..147..335D}
{Diolaiti}, E., {Bendinelli}, O., {Bonaccini}, D., {et~al.} 2000, \aaps, 147,
  335

\bibitem[{{Eckart} \& {Duhoux}(1990)}]{1990ASPC...14..336E}
{Eckart}, A. \& {Duhoux}, P.~R.~M. 1990, in ASP Conf. Ser. 14: Astrophysics
  with Infrared Arrays, ed. R.~{Elston}, 336--+

\bibitem[{{Elsasser} \& {Staude}(1978)}]{1978A&A....70L...3E}
{Elsasser}, H. \& {Staude}, H.~J. 1978, \aap, 70, L3+

\bibitem[{{Hales} {et~al.}(2006){Hales}, {Gledhill}, {Barlow}, \&
  {Lowe}}]{2006MNRAS.365.1348H}
{Hales}, A.~S., {Gledhill}, T.~M., {Barlow}, M.~J., \& {Lowe}, K.~T.~E. 2006,
  \mnras, 365, 1348

\bibitem[{{Hashimoto} {et~al.}(2007){Hashimoto}, {Tamura}, {Suto}, {Abe},
  {Ishii}, {Kudo}, \& {Mayama}}]{2007PASJ...59..221H}
{Hashimoto}, J., {Tamura}, M., {Suto}, H., {et~al.} 2007, \pasj, 59, 221

\bibitem[{{Henney}(2000)}]{2000RMxAC...9..198H}
{Henney}, W.~J. 2000, in Revista Mexicana de Astronomia y Astrofisica
  Conference Series, ed. S.~J. {Arthur}, N.~S. {Brickhouse}, \& J.~{Franco},
  198--200

\bibitem[{{Henney} \& {O'Dell}(1999)}]{1999AJ....118.2350H}
{Henney}, W.~J. \& {O'Dell}, C.~R. 1999, \aj, 118, 2350

\bibitem[{{Hester} \& {Desch}(2005)}]{2005ASPC..341..107H}
{Hester}, J.~J. \& {Desch}, S.~J. 2005, in ASP Conf. Ser. 341: Chondrites and
  the Protoplanetary Disk, ed. A.~N. {Krot}, E.~R.~D. {Scott}, \&
  B.~{Reipurth}, 107--+

\bibitem[{{Hillenbrand} {et~al.}(1998){Hillenbrand}, {Strom}, {Calvet},
  {Merrill}, {Gatley}, {Makidon}, {Meyer}, \&
  {Skrutskie}}]{1998AJ....116.1816H}
{Hillenbrand}, L.~A., {Strom}, S.~E., {Calvet}, N., {et~al.} 1998, \aj, 116,
  1816

\bibitem[{{Hughes} {et~al.}(2008){Hughes}, {Wilner}, {Qi}, \&
  {Hogerheijde}}]{2008ApJ}
{Hughes}, A.~M., {Wilner}, D.~J., {Qi}, C., \& {Hogerheijde}, M.~R. 2008,
  accepted for publication in ApJ

\bibitem[{{Jiang} {et~al.}(2005){Jiang}, {Tamura}, {Fukagawa}, {Hough},
  {Lucas}, {Suto}, {Ishii}, \& {Yang}}]{2005Natur.437..112J}
{Jiang}, Z., {Tamura}, M., {Fukagawa}, M., {et~al.} 2005, \nat, 437, 112

\bibitem[{{Johnstone} {et~al.}(1998){Johnstone}, {Hollenbach}, \&
  {Bally}}]{1998ApJ...499..758J}
{Johnstone}, D., {Hollenbach}, D., \& {Bally}, J. 1998, \apj, 499, 758

\bibitem[{{Kim} {et~al.}(1994){Kim}, {Martin}, \&
  {Hendry}}]{1994ApJ...422..164K}
{Kim}, S.-H., {Martin}, P.~G., \& {Hendry}, P.~D. 1994, \apj, 422, 164

\bibitem[{{Kuhn} {et~al.}(2001){Kuhn}, {Potter}, \&
  {Parise}}]{2001ApJ...553L.189K}
{Kuhn}, J.~R., {Potter}, D., \& {Parise}, B. 2001, \apjl, 553, L189

\bibitem[{{Lada} {et~al.}(2000){Lada}, {Muench}, {Haisch}, {Lada}, {Alves},
  {Tollestrup}, \& {Willner}}]{2000AJ....120.3162L}
{Lada}, C.~J., {Muench}, A.~A., {Haisch}, Jr., K.~E., {et~al.} 2000, \aj, 120,
  3162

\bibitem[{{Lada} {et~al.}(2004){Lada}, {Muench}, {Lada}, \&
  {Alves}}]{2004AJ....128.1254L}
{Lada}, C.~J., {Muench}, A.~A., {Lada}, E.~A., \& {Alves}, J.~F. 2004, \aj,
  128, 1254

\bibitem[{{Lant{\'e}ri} {et~al.}(1999){Lant{\'e}ri}, {Soummer}, \&
  {Aime}}]{1999A&AS..140..235L}
{Lant{\'e}ri}, H., {Soummer}, R., \& {Aime}, C. 1999, \aaps, 140, 235

\bibitem[{{Lucy}(1974)}]{1974AJ.....79..745L}
{Lucy}, L.~B. 1974, \aj, 79, 745

\bibitem[{{Mathis} {et~al.}(1977){Mathis}, {Rumpl}, \&
  {Nordsieck}}]{1977ApJ...217..425M}
{Mathis}, J.~S., {Rumpl}, W., \& {Nordsieck}, K.~H. 1977, \apj, 217, 425

\bibitem[{{Mathis} \& {Wallenhorst}(1981)}]{1981ApJ...244..483M}
{Mathis}, J.~S. \& {Wallenhorst}, S.~G. 1981, \apj, 244, 483

\bibitem[{{McCaughrean} \& {Stauffer}(1994)}]{1994AJ....108.1382M}
{McCaughrean}, M.~J. \& {Stauffer}, J.~R. 1994, \aj, 108, 1382

\bibitem[{{Meyer} {et~al.}(2006){Meyer}, {Eckart}, {Sch{\"o}del}, {Duschl},
  {Mu{\v z}i{\'c}}, {Dov{\v c}iak}, \& {Karas}}]{2006A&A...460...15M}
{Meyer}, L., {Eckart}, A., {Sch{\"o}del}, R., {et~al.} 2006, \aap, 460, 15

\bibitem[{{Monin} {et~al.}(2006){Monin}, {M{\'e}nard}, \&
  {Peretto}}]{2006A&A...446..201M}
{Monin}, J.-L., {M{\'e}nard}, F., \& {Peretto}, N. 2006, \aap, 446, 201

\bibitem[{{O'Dell}(2001)}]{2001ARA&A..39...99O}
{O'Dell}, C.~R. 2001, \araa, 39, 99

\bibitem[{{O'Dell} \& {Wen}(1994)}]{1994ApJ...436..194O}
{O'Dell}, C.~R. \& {Wen}, Z. 1994, \apj, 436, 194

\bibitem[{{O'Dell} {et~al.}(1993){O'Dell}, {Wen}, \&
  {Hu}}]{1993ApJ...410..696O}
{O'Dell}, C.~R., {Wen}, Z., \& {Hu}, X. 1993, \apj, 410, 696

\bibitem[{{Olczak} {et~al.}(2006){Olczak}, {Pfalzner}, \&
  {Spurzem}}]{2006ApJ...642.1140O}
{Olczak}, C., {Pfalzner}, S., \& {Spurzem}, R. 2006, \apj, 642, 1140

\bibitem[{{Ott} {et~al.}(1999){Ott}, {Eckart}, \&
  {Genzel}}]{1999ApJ...523..248O}
{Ott}, T., {Eckart}, A., \& {Genzel}, R. 1999, \apj, 523, 248

\bibitem[{{Pendleton} {et~al.}(1990){Pendleton}, {Tielens}, \&
  {Werner}}]{1990ApJ...349..107P}
{Pendleton}, Y.~J., {Tielens}, A.~G.~G.~M., \& {Werner}, M.~W. 1990, \apj, 349,
  107

\bibitem[{{Pruksch} \& {Fleischmann}(1998)}]{1998ASPC..145..496P}
{Pruksch}, M. \& {Fleischmann}, F. 1998, in Astronomical Society of the Pacific
  Conference Series, Vol. 145, Astronomical Data Analysis Software and Systems
  VII, ed. R.~{Albrecht}, R.~N. {Hook}, \& H.~A. {Bushouse}, 496--+

\bibitem[{{Robberto} {et~al.}(2002){Robberto}, {Beckwith}, \&
  {Panagia}}]{2002ApJ...578..897R}
{Robberto}, M., {Beckwith}, S.~V.~W., \& {Panagia}, N. 2002, \apj, 578, 897

\bibitem[{{Shuping} {et~al.}(2003){Shuping}, {Bally}, {Morris}, \&
  {Throop}}]{2003ApJ...587L.109S}
{Shuping}, R.~Y., {Bally}, J., {Morris}, M., \& {Throop}, H. 2003, \apjl, 587,
  L109

\bibitem[{{Shuping} {et~al.}(2006){Shuping}, {Kassis}, {Morris}, {Smith}, \&
  {Bally}}]{2006ApJ...644L..71S}
{Shuping}, R.~Y., {Kassis}, M., {Morris}, M., {Smith}, N., \& {Bally}, J. 2006,
  \apjl, 644, L71

\bibitem[{{Simpson} {et~al.}(2006){Simpson}, {Colgan}, {Erickson}, {Burton}, \&
  {Schultz}}]{2006ApJ...642..339S}
{Simpson}, J.~P., {Colgan}, S.~W.~J., {Erickson}, E.~F., {Burton}, M.~G., \&
  {Schultz}, A.~S.~B. 2006, \apj, 642, 339

\bibitem[{{Stauffer} {et~al.}(1994){Stauffer}, {Prosser}, {Hartmann}, \&
  {McCaughrean}}]{1994AJ....108.1375S}
{Stauffer}, J.~R., {Prosser}, C.~F., {Hartmann}, L., \& {McCaughrean}, M.~J.
  1994, \aj, 108, 1375

\bibitem[{{St{\"o}rzer} \& {Hollenbach}(1999)}]{1999ApJ...515..669S}
{St{\"o}rzer}, H. \& {Hollenbach}, D. 1999, \apj, 515, 669

\bibitem[{{Tamura} {et~al.}(2006){Tamura}, {Kandori}, {Kusakabe}, {Nakajima},
  {Hashimoto}, {Nagashima}, {Nagata}, {Nagayama}, {Kimura}, {Yamamoto},
  {Hough}, {Lucas}, {Chrysostomou}, \& {Bailey}}]{2006ApJ...649L..29T}
{Tamura}, M., {Kandori}, R., {Kusakabe}, N., {et~al.} 2006, \apjl, 649, L29

\bibitem[{{Terebey} {et~al.}(1984){Terebey}, {Shu}, \&
  {Cassen}}]{1984ApJ...286..529T}
{Terebey}, S., {Shu}, F.~H., \& {Cassen}, P. 1984, \apj, 286, 529

\bibitem[{{Throop} {et~al.}(2001){Throop}, {Bally}, {Esposito}, \&
  {McCaughrean}}]{2001Sci...292.1686T}
{Throop}, H.~B., {Bally}, J., {Esposito}, L.~W., \& {McCaughrean}, M.~J. 2001,
  Science, 292, 1686

\bibitem[{{Vicente} \& {Alves}(2005)}]{2005A&A...441..195V}
{Vicente}, S.~M. \& {Alves}, J. 2005, \aap, 441, 195

\bibitem[{{Whitney} \& {Hartmann}(1992)}]{1992ApJ...395..529W}
{Whitney}, B.~A. \& {Hartmann}, L. 1992, \apj, 395, 529

\bibitem[{{Whitney} \& {Hartmann}(1993)}]{1993ApJ...402..605W}
{Whitney}, B.~A. \& {Hartmann}, L. 1993, \apj, 402, 605

\bibitem[{{Whitney} {et~al.}(1997){Whitney}, {Kenyon}, \&
  {Gomez}}]{1997ApJ...485..703W}
{Whitney}, B.~A., {Kenyon}, S.~J., \& {Gomez}, M. 1997, \apj, 485, 703

\bibitem[{{Whitney} {et~al.}(2003){Whitney}, {Wood}, {Bjorkman}, \&
  {Cohen}}]{2003ApJ...598.1079W}
{Whitney}, B.~A., {Wood}, K., {Bjorkman}, J.~E., \& {Cohen}, M. 2003, \apj,
  598, 1079

\end{thebibliography}

\Online
\begin{appendix}

\onlfig{12}{
\begin{figure*}
\centering
\includegraphics[height=\textwidth,angle=90]{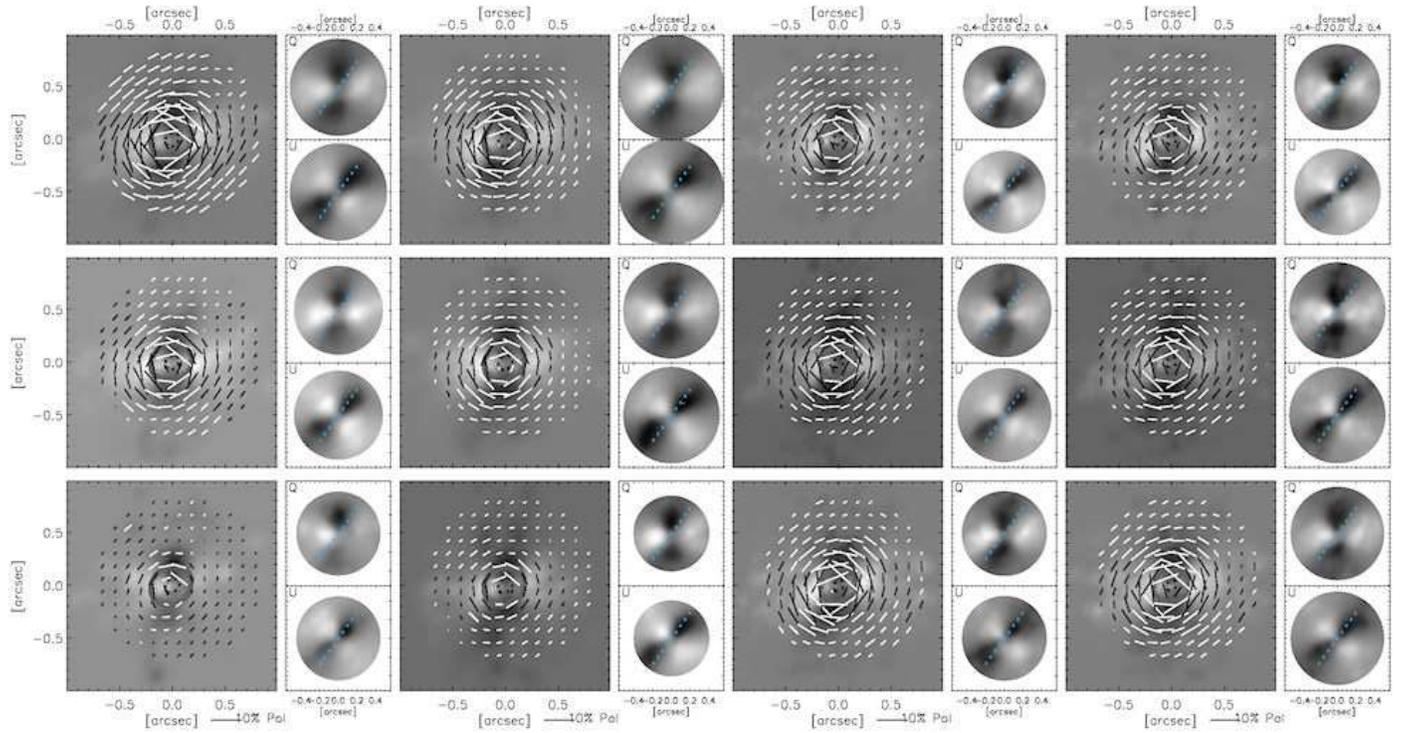}
 \caption{Comparison of the different grain models used in the simulations.
From left to right: MRN, KMH, KMH$_{DC}$ and KMH$_{DC} +$ large grains (models explained in the text)
in the disk region (each in J-, H-, and K-bands from top
to bottom). The polarization vectors and the extent of the Q- and U-patterns are shown, determined as described in 
Sect. \ref{polprop}. The images have been convolved with the PSF of the observations. 
The deviation from circular symmetry of the vectors in the outer regions is an effect of the elongated PSF.}
\label{grains}
\end{figure*} 
}

\onlfig{13}{
\begin{figure*}
\centering
\includegraphics[width=\textwidth]{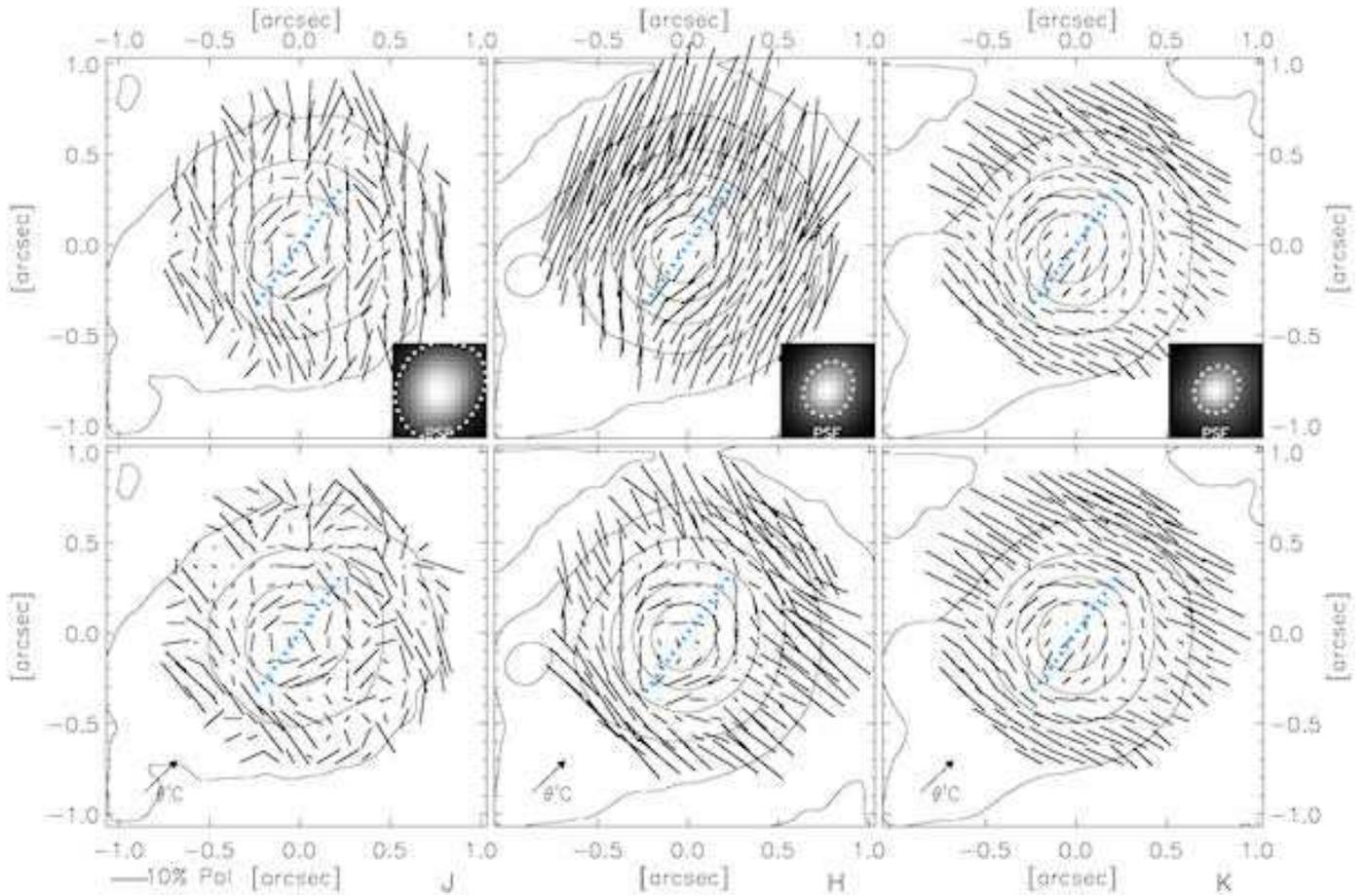}
 \caption{Polarization vector maps before (top) and after shifting the Q- and U-images (bottom). The
offset is determined from the asymmetry around zero for the azimuthal Q- and U-profiles close to the star (0.35'' in J-band). The shifting
compensates an offset in polarized flux with fixed angle in the corresponding aperture 
induced by the elongation of the respective field PSFs in the direction of $\theta^1$ Ori C.
The offset dominates in the outer regions, where the polarization degree is high, but the total polarized flux is low.
As expected, the effect is irrelevant close to the star. However, in 
H-band (middle) the effect is very strong for the reflection nebula 
and is interpreted to be dominated by errors in the polarimetric calibration process, as the flux
in the outer regions is close to the noise and instrumental polarization. Hence, the shifting based on the higher flux close to the star
provides a bootstrap calibration since the obtained vector pattern is what we expect for the reflection nebula. 
}
\label{shiftvectors}
\end{figure*} 
}
\onlfig{14}{
\begin{figure*}
\centering
\includegraphics[width=\textwidth]{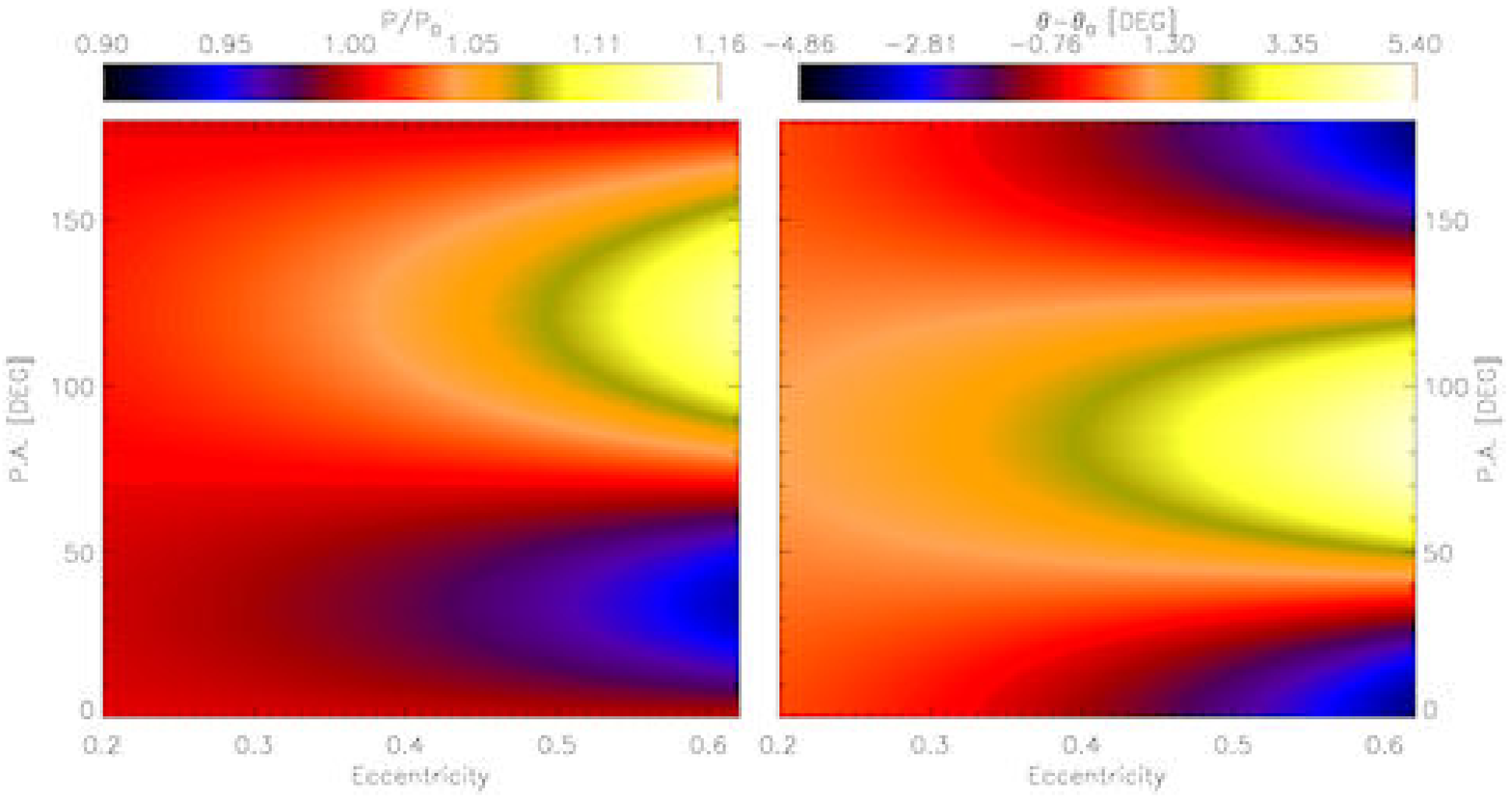}
\includegraphics[width=\textwidth]{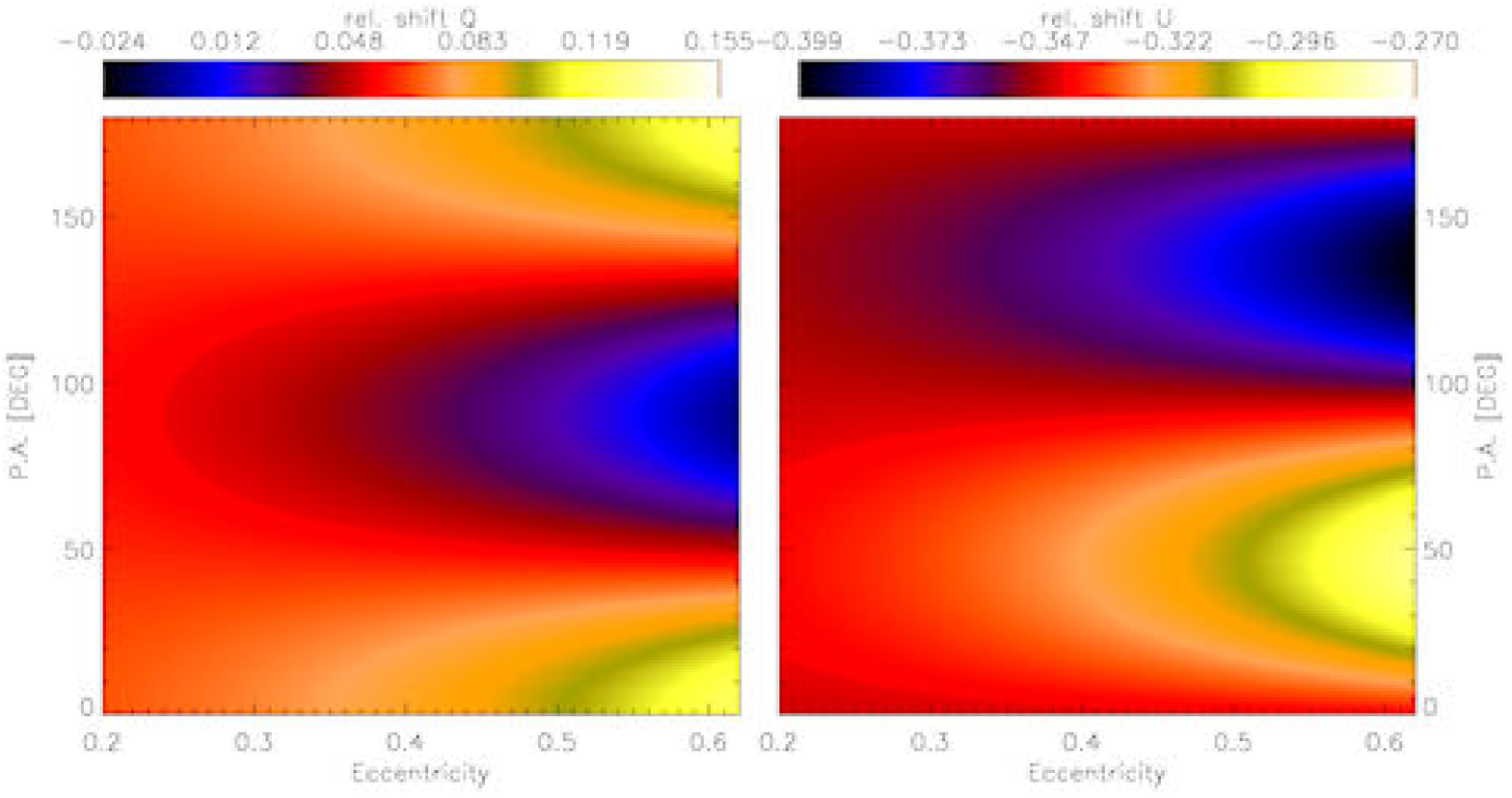}
 \caption{Effects of convolution with an elongated PSF on polarization. The model is simulated in H-band and 
includes a flat disk with moderate mass ($M=0.05M_{\odot}$, $h=0.001R_S$) and an inclination of $\xi=75^\circ$.
The images were convolved with artificial PSFs of 0.3'' FWHM with varying eccentricities and
rotated from 0$^\circ$ to 180$^\circ$. Top: integrated polarization degree and angle 
in an circular aperture of 0.3''. 
The values are given in relation to the results for a circular symmetric PSF ($P_0$, $\theta_0$). Bottom:
shift of the Q- and U-images relative to the maximum amplitude of the profile. }
\label{elongeffects}
\end{figure*} 
}
\onlfig{15}{
\begin{figure*}
\centering
\includegraphics[width=\textwidth]{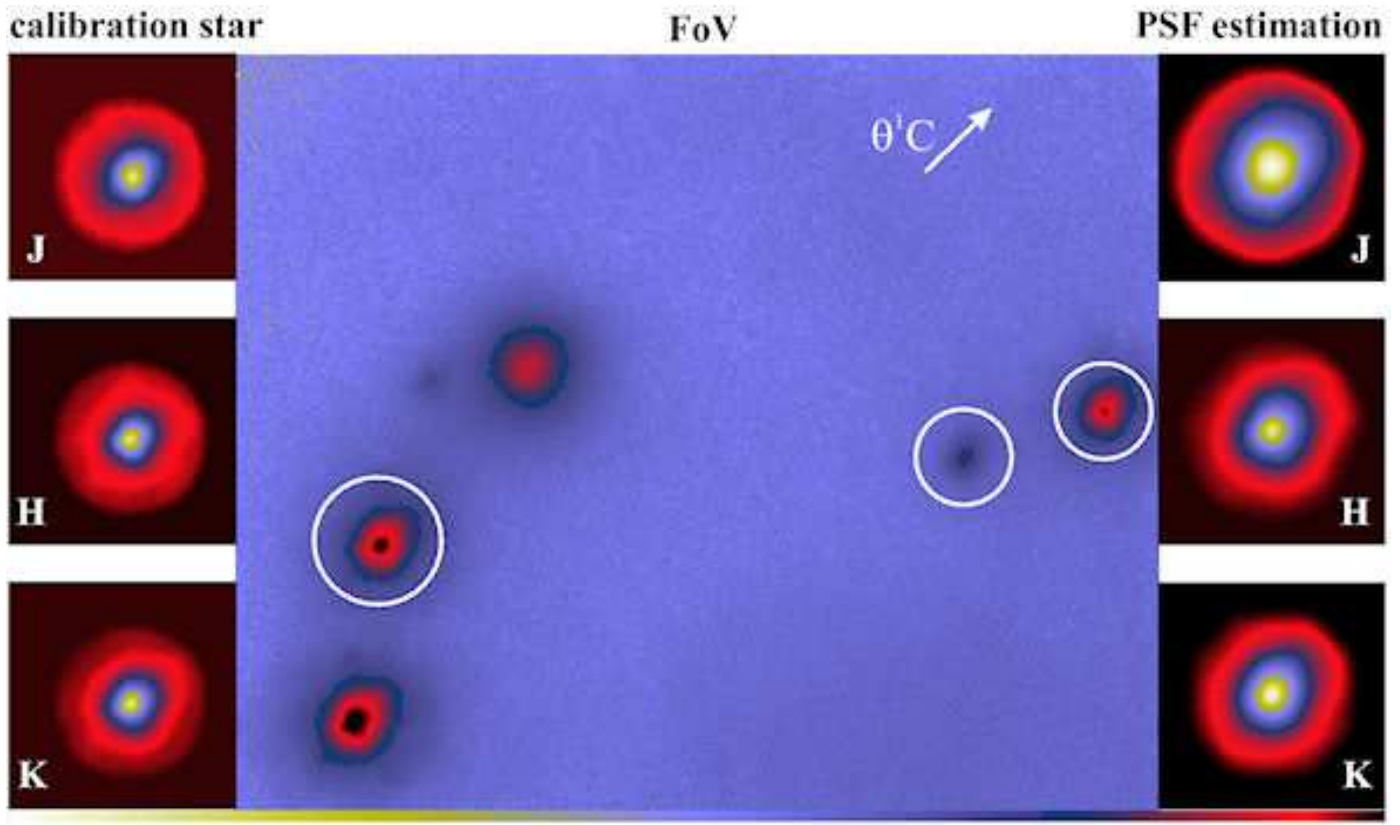}
 \caption{PSF estimation from the FOV in comparison to the calibration star PSFs. Three point-like sources are chosen to
extract the PSF from the field. The elongation is higher in the field PSFs, the performance in J-band is much lower.
The eccentricities obtained by fitting the field-PSFs are 0.27, 0.35 and 0.26 for J-, H- and K-band.}
\label{psfestim}
\end{figure*} 
}
\end{appendix}


\end{document}